\documentclass{article}

\usepackage{arxiv}

\usepackage[utf8]{inputenc} 
\usepackage[T1]{fontenc}    
\usepackage{hyperref}       
\usepackage{url}            
\usepackage{booktabs}       
\usepackage{amsfonts}       
\usepackage{nicefrac}       
\usepackage{microtype}      
\usepackage{lipsum}

\usepackage{rotating}
\usepackage{textcomp}
\usepackage{amsmath}
\usepackage{threeparttable}

\title{A Model-View-ViewModel (MVVM) Application Framework for Hearing Impairment Diagnosis}

\author{
  Waseem Sheikh \\
  CSET Department \\
  Oregon Institute of Technology \\
  Klamath Falls, OR 97601 \\
  \texttt{waseem.sheikh@oit.edu} \\
   \And
  Nadeem Sheikh \\
  Department of ENT \\
  Combined Military Hospital\\
  Abbottabad, Pakistan \\
  \texttt{shingoveteran@hotmail.com} \\
}

\begin{document}
\maketitle

\begin{abstract}
Around 466 million people worldwide (over 5\% of the world's population) have disabling hearing loss, and out of these 34 million are children. Estimates suggest that by 2050, over 900 million people worldwide will have disabling hearing loss. The annual global cost of unaddressed hearing loss amounts to US\$ 750 billion. Early detection of hearing loss can reduce its impact on an individual's life in addition to saving a huge cost.

This paper presents the design, implementation, and evaluation of an open-source application framework for hearing impairment diagnosis. The framework is built using the Model-View-ViewModel (MVVM) pattern which separates the development of graphical user interface (GUI) from the development of business and back-end logic. Some of the benefits of the MVVM pattern include reusable components, independent development of GUI and business or back-end logic, flexibility to modify GUI without having to change business or back-end logic, ease of unit testing, and reduced maintenance overhead. The proposed framework along with the open-source code makes it possible to easily extend the application functionality thus enabling other researchers and practitioners to develop their own versions of hearing loss diagnosis applications. The proposed software was evaluated by an otolaryngologist and found to be very beneficial in assisting a clinician to reach a hearing impairment diagnosis conclusion more methodically, swiftly and accurately.
\end{abstract}


\section{Introduction}\label{sec-intro}

According to a recent World Health Organization (WHO) report \cite{Who2019}, around 466 million people worldwide (over 5\% of the world's population) have disabling hearing loss, and out of these 34 million are children. The report estimates that by 2050 over 900 million people worldwide - one in every ten people - will have disabling hearing loss. In the United States, approximately 15\% of people over the age of 18 years (37.5 million) report some level of hearing loss.\cite{SummaryHealthStatistics2017,Blackwell2014} Low- and middle-income countries have the highest proportion of people with disabling hearing loss.\cite{Who2019} 

A hearing threshold above 25 dB in at least three speech frequencies (500 Hz, 1 kHz, and 2 kHz) in either ear is defined as a hearing loss or hearing impairment.\cite{KramerAudiology2019,GelfandEssentials2016,KatzHandbook2015,DhingraDiseases2014,BessAudiology2003} WHO defines disabling hearing loss as a hearing loss greater than 40 dB in the better hearing ear in adults and a hearing loss greater than 30 dB in the better hearing ear in children.\cite{Who2019} In the United States, disabling hearing impairment is defined by the Individuals with Disabilities Education Act (IDEA) as ``an impairment in hearing, whether permanent or fluctuating, that adversely affects a child's educational performance but is not included under the definition of `deafness'.'' IDEA defines deafness as ``a hearing impairment that is so severe that the child is impaired in processing linguistic information through hearing, with or without amplification.'' \cite{Idea2019,KnoblauchIdea1998}

Undetected hearing loss has significant negative implications on the socioeconomic status of an individual and the societal costs. It impacts an individual's ability to communicate with others. This impact is far worse on children whose academic performance suffers. Lack of ability to communicate can result in feelings of loneliness and isolation. The WHO report estimates that the annual global cost of unaddressed hearing loss amounts to around US\$ 750 billion, which includes health sector costs, costs of educational support, loss of productivity, and societal costs.\cite{Who2019} In developing countries, children with hearing loss are not able to finish their schooling. Adults with hearing loss tend to have a much higher unemployment rate. Early detection of hearing loss and intervention is crucial in reducing its impact on an individual's life and reducing the societal cost.\cite{Who2019}

An ear, nose, and throat (ENT) doctor, also known as as an otolaryngologist, or an audiologist may use a variety of tests to identify and diagnose a particular kind of hearing impairment. Some of these tests include tuning fork tests such as Weber, Rinne, Schwabach, absolute bone conduction, Teal, and Gelle; speech audiometry; pure-tone audiometry (PTA); impedance audiometry; bithermal caloric test; and advanced tests such as alternate binaural loudness balance (ABLB), short increment sensitivity index (SISI), tone decay, and Stenger.\cite{KramerAudiology2019,GelfandEssentials2016,KatzHandbook2015,DhingraDiseases2014,BessAudiology2003} These hearing tests use specialized hardware including an audiometer, a tympanometer, tuning forks, and vestibular testing equipment; and software to process and visualize data in standard reproducible formats such as a pure-tone audiogram, speech audiogram, tympanogram, calorigram, and loudness balance chart. Even though proprietary hearing tests related hardware and software exist from various vendors, the price of such equipment makes it prohibitive for low- and middle-income countries which tend to have a higher prevalence of people with hearing loss. In most underdeveloped and developing countries, hearing test-related data is stored on paper and graphs such as audiograms are drawn by hand. Such a system of managing hearing test data makes it very difficult to save, track, compare and reproduce hearing test data. In addition, a lack of open-source software in this domain stifles innovation.

This paper presents the design and implementation of an open-source application framework for accurate digital recording and graphical description of human audio-vestibular impairment test data to assist in hearing loss or disability diagnosis. The framework is built using the Model-View-ViewModel (MVVM) software architectural pattern which separates the development of graphical user interface (GUI) from the development of business and back-end logic.\cite{MvvmWikipedia2019} Some of the benefits of the MVVM pattern include reusable components, independent development of GUI and business or back-end logic, flexibility to modify GUI without having to change business or back-end logic, ease of comprehensive unit testing, faster application development time, and reduced maintenance overhead.\cite{MvvmMicrosoft2019} The proposed framework along with the open-source code makes it possible to easily extend the application functionality thus enabling other researchers and practitioners to develop their own hearing impairment diagnosis-related applications. In addition, the proposed framework is not restricted to the specialized domain of hearing impairment diagnosis but can be easily extended to other medical testing domains.

The main contributions of the paper along with the open-source software are:
\begin{itemize}
\item  An open-source hearing test application that can store, process, and visualize data corresponding to tuning fork tests including Weber, Rinne, Schwabach, absolute bone conduction, Teal, and Gelle; speech audiometry; pure-tone audiometry (PTA); impedance audiometry; bithermal caloric test; and advanced tests including alternate binaural loudness balance (ABLB), short increment sensitivity index (SISI), tone decay, and Stenger.\cite{KramerAudiology2019,GelfandEssentials2016,KatzHandbook2015,DhingraDiseases2014,BessAudiology2003}
\item An open-source hearing test application framework that can be used to develop new hearing test applications by extending the current functionality. One of the main requirements for the architectural design was extensibility of the application.
\item The application framework is independent of specific hearing test hardware thus making it possible to be used with any hearing test hardware.
\item A unified and uniform interface for storing, processing, and visualizing data from a wide range of hearing tests which traditionally rely on different hardware and software to process and store data.
\item Important architectural design and lessons learnt from the implementation and testing of the application that can be useful in other domains in addition to hearing impairment diagnosis.
\item Results from an evaluation study of the software by an otolaryngologist who found the software be very beneficial in reaching a hearing impairment diagnosis conclusion more methodically, swiftly and accurately.
\end{itemize}

This paper is organized as follows: Section~\ref{sec-related-work} provides an overview of the recent research in the area of computer, tablet, and smartphone-based hearing test and screening applications; Section~\ref{sec-application-features} presents the application requirements and features implemented in the proposed hearing test application; a detailed architectural design and implementation of the proposed application framework is presented in Section~\ref{sec-design-implementation}; Section~\ref{sec-practical-experience} discusses testing and practical experience using the application; details of how to access the source code, documentation, and installation files for the application are presented in Section~\ref{sec-code-deployment}; and finally, Section~\ref{sec-conclusion} concludes the paper with possible future research directions.

\section{Related Work}\label{sec-related-work}

Recently there has been an increased interest in the areas of tele-audiology and computer, tablet, or smartphone-based hearing test and screening applications. The authors in \cite{ChenSmartphone2017,ChenSmartphone2018} propose a smartphone-based audiometry method for hearing self-assessment, which uses hearing aids as the audio sources and a smartphone as the user interface and controller. The test result is output in the form of an audiogram. The design of a Web services-based system for remote hearing diagnosis is presented in \cite{YaoWebHearingDiagnosis2009}. This distributed system follows a browser-server architecture and makes basic hearing test services more accessible to traditionally underserved areas. The system stores the audiological profile data into a standard database that can be integrated with other established electronic medical records. In \cite{KochanekAudiometer2007}, the authors present the construction and implementation of a computer-based audiometer for hearing testing and screening of newborns and young children. The device consists of a computer interfaced with an otoacoustic emission (OAE)/auditory brainstem response (ABR) measuring probe. The system enables measurement of OAE, ABR, as well as pure-tone audiogram.

A browser-server-based remote hearing test system that integrates a pure-tone audiogram and a speech test is presented in \cite{YaoBrowser2015}. This system consists of a Web application server, an embedded smart Internet-Bluetooth\textsuperscript{\textregistered} gateway (or console device), and a Bluetooth-enabled audiometer. A Microsoft\textsuperscript{\textregistered} Excel\textsuperscript{\textregistered}-based pure-tone audiometry software tool for the classification of audiograms for the inclusion of patients in clinical trials was developed in \cite{Rahne2016}. An efficacy comparison study of a Web-based hearing screening application was presented in \cite{SerenWeb2009}. The audiometric thresholds measured by the Web-based system and a conventional audiometer varied by no more than 1.78 dB for air conduction thus showing the feasibility of the Web-based system. A pilot study of an Internet-based tele-audiometric system for hearing assessment is presented in \cite{GivensInternet2003}. The auditory threshold data from the Internet-based and conventional systems show substantial agreement. A Windows\textsuperscript{\textregistered} software application for audiogram analysis and hearing research is presented in \cite{JobVisaudio1996}. The application provides graphical displays of audiograms with the results of various processing such as smoothing, pathological patterns detection, and classification. The software also performs audiometric pattern classification for each audiogram using discriminant factorial analysis and allows comparisons of audiograms.

The researchers in \cite{FouladAutomated2013} present a feasibility study of an Apple\textsuperscript{\textregistered} iOS\textsuperscript{\textregistered}-based automated audiometry application. The application performed pure-tone hearing test on the iPhone\textsuperscript{\textregistered}, iPod\textsuperscript{\textregistered} touch, and iPad\textsuperscript{\textregistered}. The application yielded hearing test results that approach that of conventional audiometry. A tablet-based tele-audiometry method for automated hearing screening of schoolchildren is described in \cite{SamelliTablet2018,SamelliTablet2017}. The performance of the tablet-based audiometry was assessed by comparing its results with the gold-standard pure-tone audiometry. The results showed that the tablet-based hearing screening test was a reliable and accurate method for hearing screening.

Mixed results have been obtained regarding the validity of smartphone-based hearing screening applications (apps) with some studies showing a significant difference between smartphone apps and conventional audiometry while others showing the two to be more in agreement. Some of these studies have found that even though smartphone apps can be used as a highly accessible portable audiometer, these apps may not be able to accurately determine the level of hearing impairment when compared to a standard audiometer administered by a trained personnel.\cite{AbuSmartphone2016,LivshitzApplication2017,BarczikAccuracy2018,HandzelSmartphone2013,McphersonHearing2010} On the other hand, some studies found no significant difference between smartphone-based hearing screening apps and conventional audiometry in terms of sensitivity and specificity.\cite{MahomedClinical2016,RendaSmartphone2016} 

Even though there has been a considerable research activity, in the recent years, in the areas of tele-audiology and computer, tablet, or smartphone-based portable hearing screening and test applications, these contributions have the following shortcomings which are addressed by our contribution:
\begin{itemize}
\item  The current research is limited to the evaluation of proprietary audiometry software and hardware. These software applications are not extensible as the source code is not freely available. We, on the other hand, provide an open-source hearing test application framework that can be easily extended to develop new hearing test, screening, and diagnosis applications.
\item  Most of the computer-based hearing test applications concentrate on pure-tone air conduction audiometry. There are numerous other tests that are employed by an otolaryngologist or an audiologist to identify and diagnose a particular kind of hearing impairment such as tuning fork tests, free field and speech audiometry, tonal and impedance audiometry, and bithermal caloric testing.\cite{KramerAudiology2019,GelfandEssentials2016,KatzHandbook2015,DhingraDiseases2014,BessAudiology2003} Our proposed application framework provides a mechanism to store, process, and visualize test data for all of the aforementioned tests. In addition, the framework can be easily extended to incorporate additional hearing tests.
\item In most of the existing computer-based audiometry applications, either the audiograms are incomplete or do not fully conform to the American National Standards Institute (ANSI) ANSI S3.6-1996 Specification for Audiometers standard.\cite{ANSIS3.6-1996} The audiogram generated as part of our application is complete and fully conforms to the ANSI S3.6-1996 standard.\cite{ANSIS3.6-1996}
\item The prevalent hearing test and diagnosis software is proprietary, closed source, and tightly coupled with the specific vendor hardware. Our contribution provides an open-source general framework for hearing test and diagnosis applications, which is not coupled with any particular vendor hardware but can rather be used with any existing hearing test hardware. 
\item As mentioned above, some of the research points to the limitations of tablet or smartphone-based audiometry applications. By decoupling the software from the particular hardware, the accuracy and validity of the hearing test data are not compromised.
\item Existing suite of hearing tests use different hardware and software to store, process, and visualize data. By providing a centralized and consistent interface for hearing test data, an otolaryngologist or an audiologist will be able to perform a more accurate diagnosis in a shorter period of time.
\item The existing applications do not provide a mechanism for implementing various data analytics algorithms to draw important conclusions from the hearing test data. On the other hand, our proposed open-source application framework opens up the possibility of integrating various machine learning techniques to process test data for automatic diagnosis, prediction, and classification.
\end{itemize}

\begin{figure}[t]
\centerline{\includegraphics[scale=1.5]{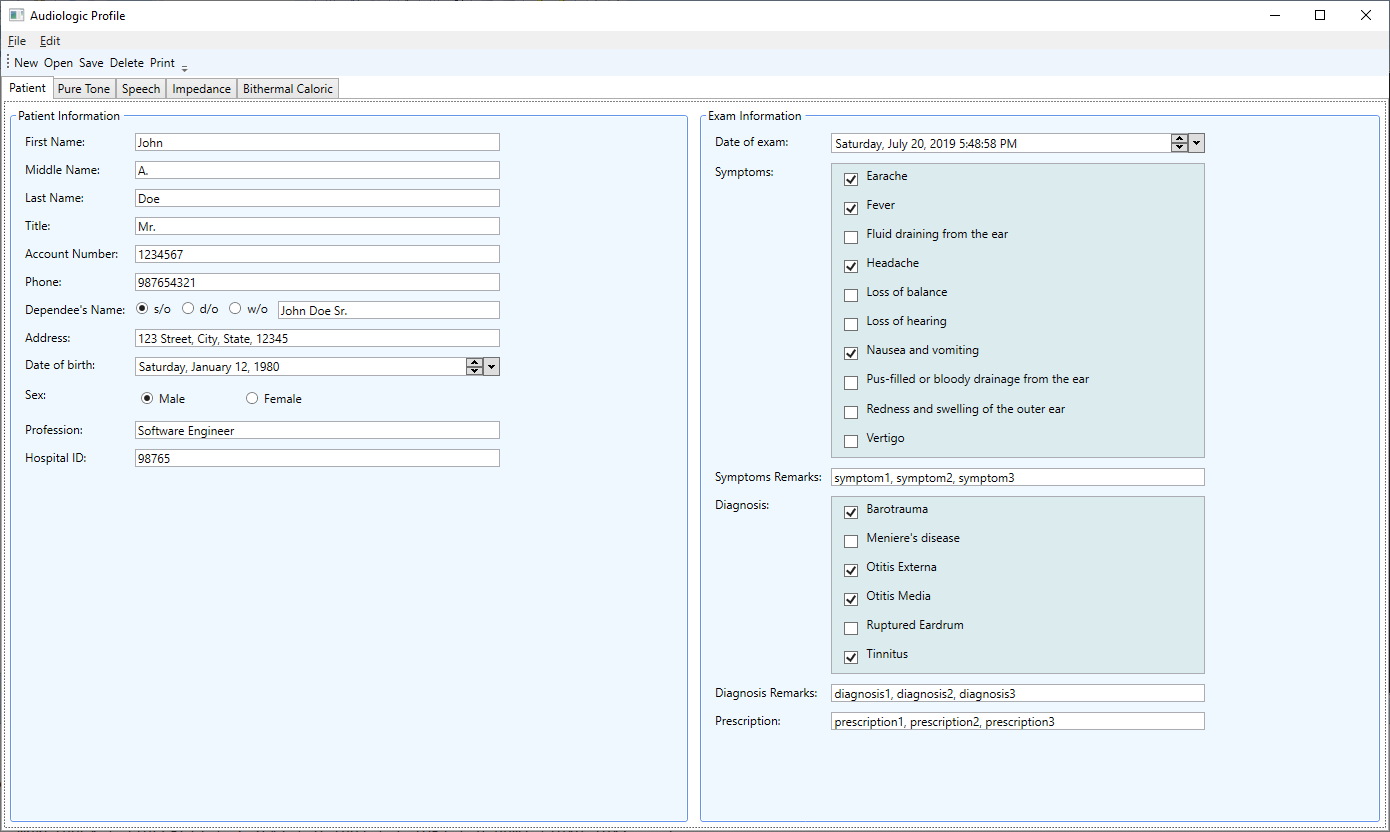}}
\caption{Patient information interface of the application.\label{fig-patient1}}
\end{figure}

\section{Application Requirements and Features}\label{sec-application-features}

In this section, we describe the main requirements and features of the application. The requirements and features were developed after extensive consultations with an otolaryngologist with many years of expertise in hearing testing and diagnosis, who is also the co-author of this paper. The expertise and insight of a domain expert were highly valuable throughout the software development process but especially during the requirements engineering phase.

At a high level, the main features of the application are to persistently store, search, edit, compute, print, export, and visualize hearing test data to help in the diagnosis of hearing impairment. The application provides interfaces for tuning fork tests including Weber, Rinne, Schwabach, absolute bone conduction, Teal, and Gelle; speech audiometry; pure-tone audiometry (PTA); impedance audiometry; bithermal caloric test; and advanced tests including alternate binaural loudness balance (ABLB), short increment sensitivity index (SISI), tone decay, and Stenger.\cite{KramerAudiology2019,GelfandEssentials2016,KatzHandbook2015,DhingraDiseases2014,BessAudiology2003} The various tests are grouped together into separate interfaces based on their functional relationship.

\subsection{Patient Interface}\label{subsec-patient-interface}

Figure~\ref{fig-patient1} shows the interface to store a patient's personal and hearing impairment diagnosis data. After the tests are conducted, and an otolaryngologist or audiologist has diagnosed the particular kind of hearing impairment, the particular diagnosis can be entered on this interface along with detailed notes for symptoms, diagnosis, and prescriptions.

\begin{figure}[t]
\centerline{\includegraphics[scale=1.5]{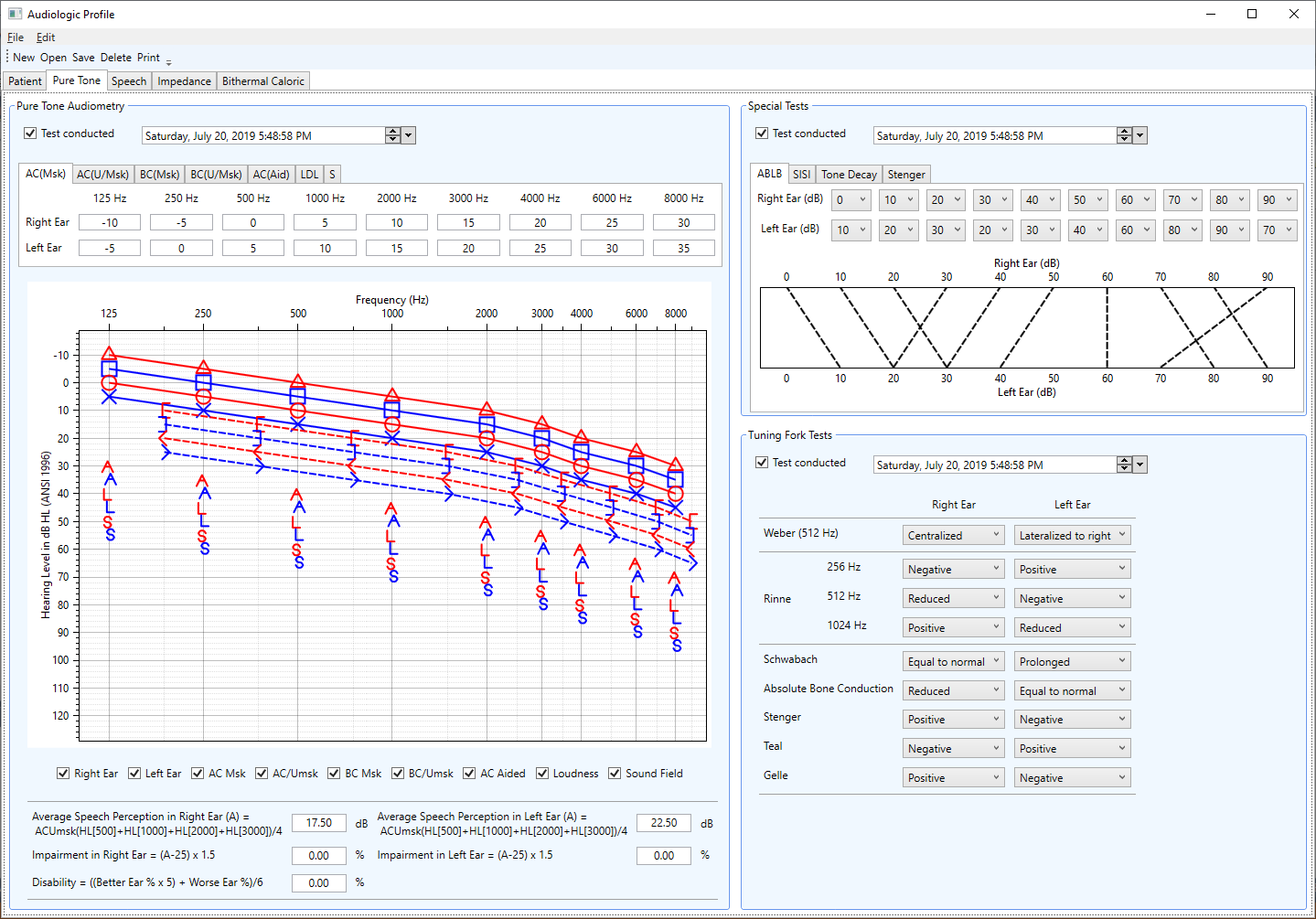}}
\caption{Pure-tone audiometry interface of the application.\label{fig-puretone1}}
\end{figure}

\begin{figure}[t]
\centerline{\includegraphics[scale=1.5]{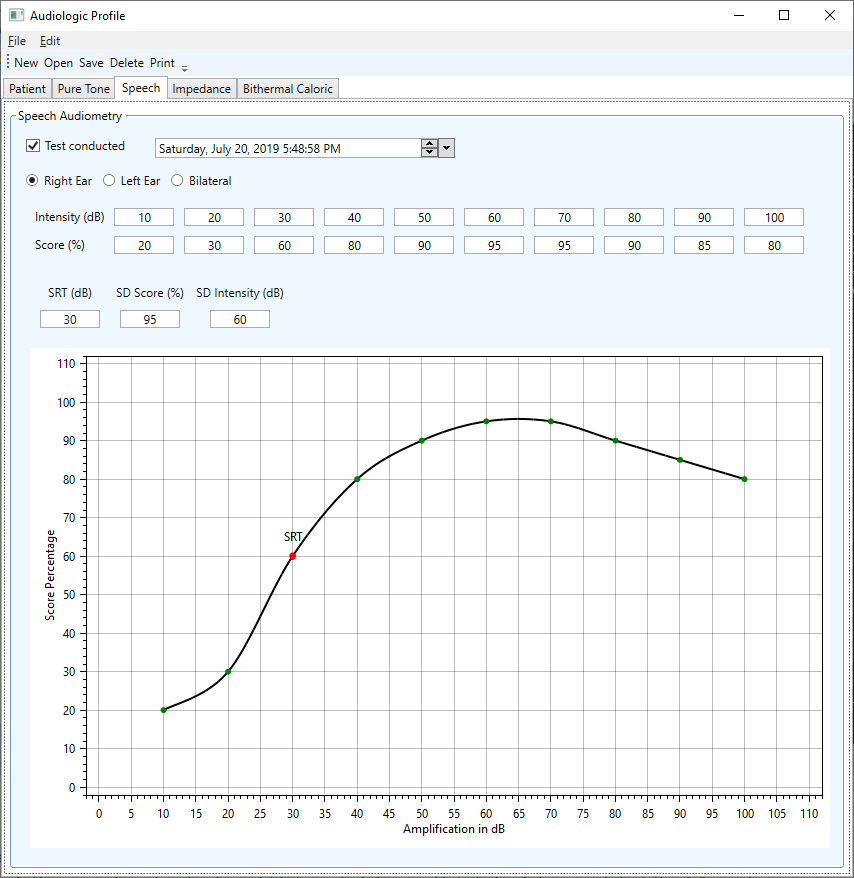}}
\caption{Speech audiometry interface of the application.\label{fig-speech1}}
\end{figure}

\subsection{Pure-Tone Audiometry (PTA) Interface}\label{subsec-puretone-interface}

The interface for pure-tone audiometry (PTA) related tests is shown in Figure~\ref{fig-puretone1}. PTA is administered to determine the hearing threshold levels of an individual to frequency-specific pure tones generally between 125 Hz to 8 kHz. This test helps in the determination of the degree, type, and configuration of a hearing loss.\cite{KramerAudiology2019,GelfandEssentials2016,KatzHandbook2015,DhingraDiseases2014,BessAudiology2003} The application interface for PTA provides the ability to store pure-tone audiometry data for air conduction (masked, unmasked, and aided), bone conduction (masked and unmasked), loudness level, and sound field. The interface draws the associated pure-tone audiogram as the audiometry data is entered. A pure-tone audiogram is a graph that shows the audible threshold for standardized pure-tone speech frequencies as measured by an audiometer. The generated audiogram plot conforms to the ANSI S3.6-1996 Specification in terms of the colors, marker types, and line shapes.\cite{ANSIS3.6-1996} By observing the shape of the curves in the audiogram plot for various combinations of air conduction (masked, unmasked, and aided), bone conduction (masked and unmasked), loudness level, and sound field; an otolaryngologist or an audiologist is able to determine the degree, type, and configuration of a hearing loss. The audiogram interface provides the ability to show any combination of the curves for air conduction (masked, unmasked, and aided), bone conduction (masked and unmasked), loudness level, and sound field, for either right or left ears. This flexibility in the visualization of pure-tone audiogram enhances the ability of an otolaryngologist or audiologist to diagnose the particular kind of hearing loss.

The interface also computes various metrics to determine hearing disability. These metrics include average speech perception and hearing impairment for both right and left ears, and hearing disability. These metrics are displayed on the user interface after the pure-tone audiometry data is entered and are computed as follows:\cite{Bauman2016}

\begin{eqnarray}
\text{Average speech perception (A)}  &= &\frac{\text{HL}[500]+\text{HL}[1000]+\text{HL}[2000]+\text{HL}[3000]}{4}\  (\text{dB}) \label{eq-perception} \\
\text{Hearing impairment} &= &\text{max}\{ (\text{A}-25) \times 1.5, 0 \}\  (\%) \label{eq-impairment} \\
\text{Hearing disability} &= &\frac{[(\text{Better Ear }\% \times 5)+\text{Worse Ear }\% ]}{6}\  (\%) \label{eq-disability}
\end{eqnarray}

where HL[500], HL[1000], HL[2000], and HL[3000] in equation~\ref{eq-perception} are the unmasked air conduction audiometry threshold levels in dB HL for 500 Hz, 1 kHz, 2 kHz, and 3 kHz pure-tone frequencies, respectively. Equations~\ref{eq-perception} and \ref{eq-impairment} are evaluated for both right and left ears. In equation~\ref{eq-disability}, Better Ear \% refers to the smaller of the hearing impairment levels of the two ears as computed in equation~\ref{eq-impairment}.

The other interfaces on the PTA tab consist of a collection of special tests and tuning fork tests. The special tests section includes interfaces for ABLB, SISI, tone decay, and Stenger. The ABLB test determines a comparison of the intensity levels in dB hearing level (HL) at which a given pure-tone sounds equally loud to the normal ear and to the ear with hearing loss.\cite{GelfandEssentials2016} A loudness balance chart or laddergram is drawn as part of the ABLB interface.\cite{GelfandEssentials2016} The SISI test measures the ear's ability to detect small intensity changes. In this test, twenty short pulses of 1 dB intensity, 200-millisecond duration, and spaced at 5-seconds intervals are superimposed onto a carrier tone of 20 dB sensation level (SL). SL is the number of decibels above the hearing threshold. The patient listens to the carrier tone and indicates when the brief pulses are heard. The test is scored by finding the percentage of twenty pulses that were heard. The SISI test is used to determine if the patient has cochlear pathology.\cite{GelfandEssentials2016} In tone decay test, a tone of 4 kHz at progressively increasing 5 dB SL levels is presented continually to the patient until the patient hears it for the complete duration of 60 seconds. This test is used for the diagnosis of retrocochlear lesion.\cite{DhingraDiseases2014} The Stenger test is used to test for unilateral nonorganic hearing loss. In this test, a pair of identical tones and different intensity levels are applied to the two ears and the ear which hears the tone is noted.\cite{KramerAudiology2019,KatzHandbook2015,DhingraDiseases2014} The tuning fork tests section includes interfaces for Weber, Rinne, Schwabach, absolute bone conduction, Teal, and Gelle tests. These tests are an important supplement to PTA and provide qualitative information to help determine whether the hearing loss is conductive or sensory/neural.\cite{KramerAudiology2019,KatzHandbook2015,DhingraDiseases2014}

\begin{figure}[t]
\centerline{\includegraphics[scale=1.5]{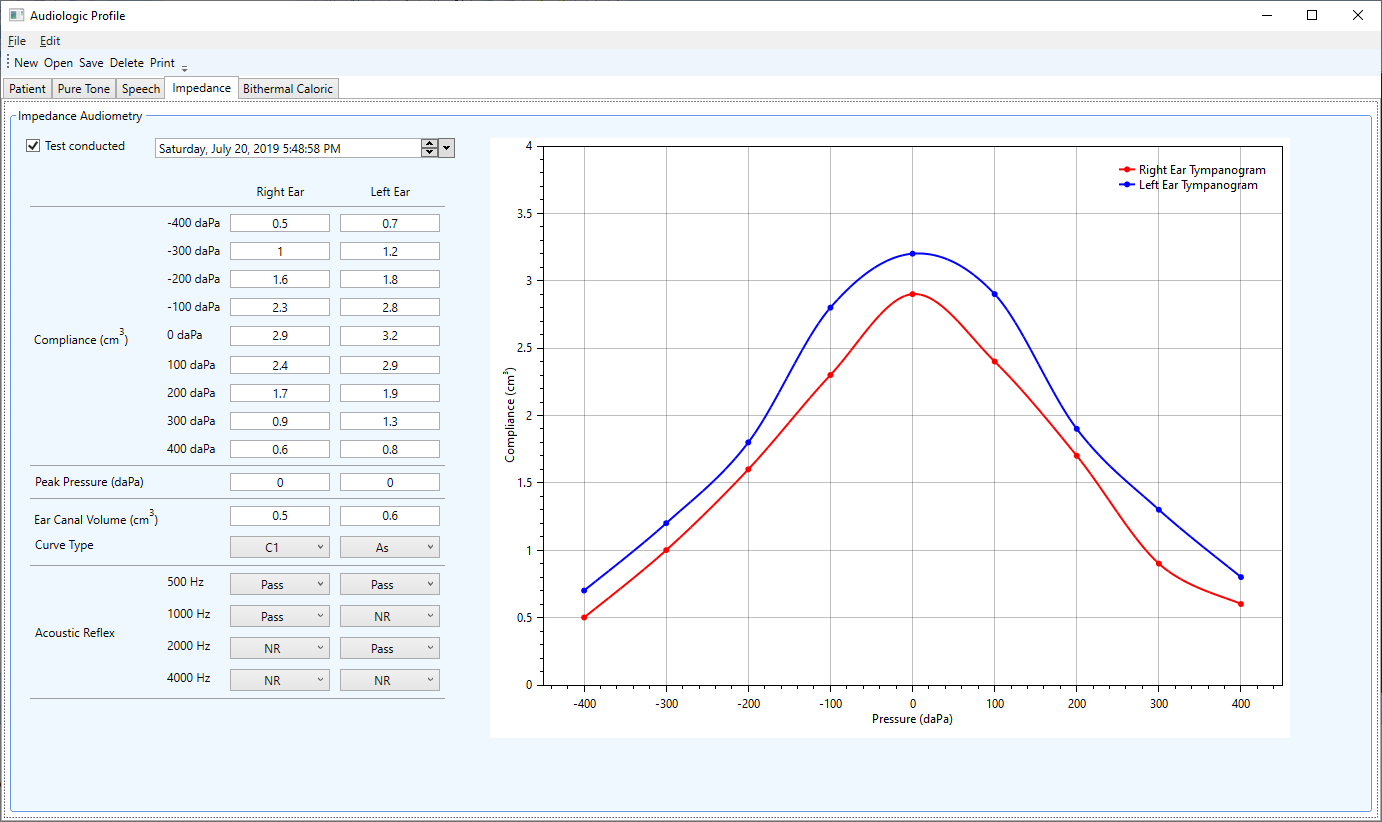}}
\caption{Impedance audiometry interface of the application.\label{fig-impedance1}}
\end{figure}

\begin{figure}[htbp]
\centerline{\includegraphics[scale=1.5]{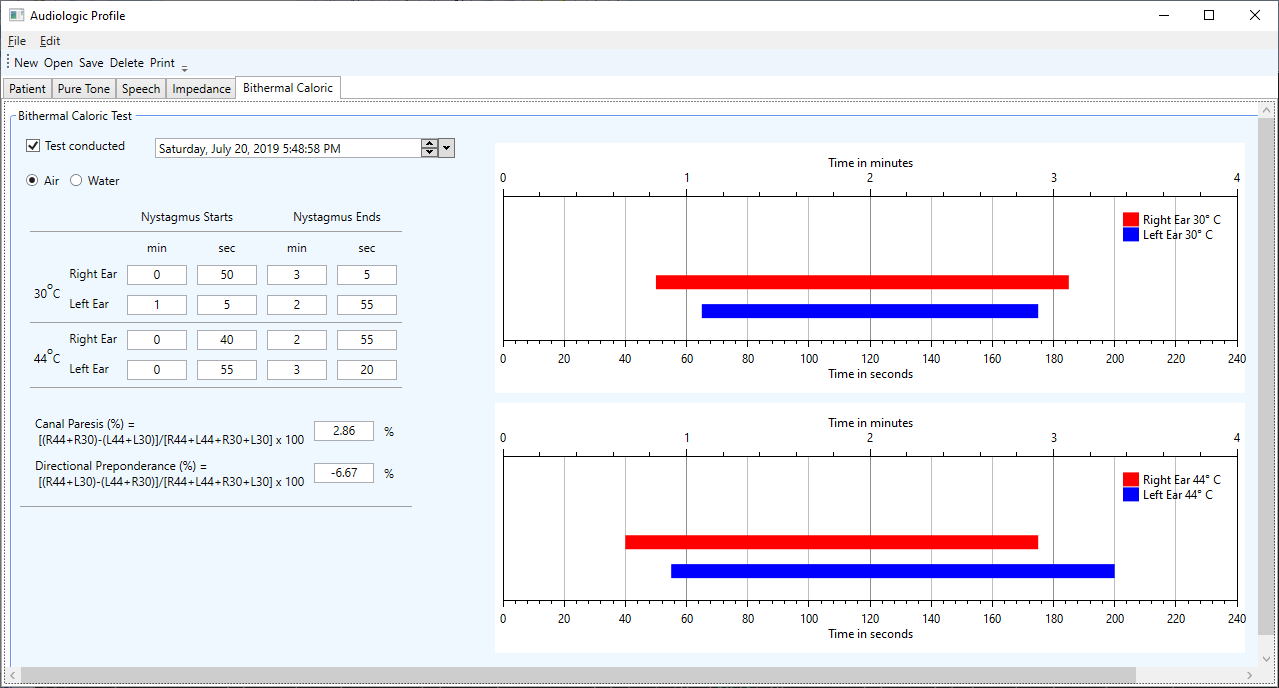}}
\caption{Bithermal caloric test interface of the application.\label{fig-calorigram1}}
\end{figure}

\subsection{Speech Audiometry Interface}\label{subsec-speech-interface}
PTA provides a partial picture of the function of the auditory system since it does not provide a precise measure of a person's ability to hear and understand speech. In speech audiometry a patient is tested with speech stimuli instead of pure-tone stimuli as in PTA. This test is performed using a speech audiometer. The outputs of the test are a speech audiogram, speech reception threshold (SRT), speech discrimination (SD) score, and SD intensity. A set of spondee words is delivered to the patient's ears through the headphones of a speech audiometer. The intensity of the word sequence is varied in steps of 5 dB until 50\% of them are correctly heard. The minimum intensity at which 50\% of the words are repeated correctly by the patient is called the SRT. To determine the SD score and intensity, a list of phonetically balanced words is delivered to each ear separately at 30-40 dB above the patient's SRT and the percentage of words correctly heard is recorded. This percentage is called the SD Score and the intensity at which it was recorded is the SD intensity. The speech audiometry interface of the application is shown in Figure~\ref{fig-speech1}.\cite{KramerAudiology2019,GelfandEssentials2016,KatzHandbook2015,DhingraDiseases2014,BessAudiology2003} The interface provides functionality to enter data collected from the speech audiometer, plot the speech audiogram, and compute SRT, SD score, and SD intensity.

\subsection{Impedance Audiometry Interface}\label{subsec-impedance-interface}
When sound wave strikes a tympanic membrane, part of the wave is transmitted through the middle ear to the cochlea, while the remaining part of the wave is reflected out of the external canal. The proportion of energy reflected back carries information regarding the condition of the middle ear system which helps in the diagnosis of conditions like otitis media and otosclerosis. A stiffer tympanic membrane reflects more energy than a complaint. Impedance is a measure of the resistance to the flow of energy and impedance audiometry measures this resistance to the flow of a sound wave. Some modern devices measure admittance which is the reciprocal of impedance and measures the ease with which sound energy flows through the ear. Collectively admittance and impedance are sometimes known as immittance.\cite{KramerAudiology2019,GelfandEssentials2016,KatzHandbook2015,DhingraDiseases2014,BessAudiology2003}

There are two main components of impedance audiometry: tympanometry and acoustic reflex measurements. In tympanometry, using specialized instrument called a tympanometer, a varying amount of pressure is applied into the airtight ear canal and a low-frequency sound wave is generated using a speaker inside the ear canal. The energy of the sound wave reflected back from the tympanic membrane is measured using a microphone also placed inside the ear canal. These measurements are repeated for different pressure values. A plot of the impedance measurements in units arbitrarily labeled as compliance versus pressure values in decapascal (daPa) is called a tympanogram. The shape of the tympanogram helps in the identification of certain middle ear pathologies.\cite{KramerAudiology2019,GelfandEssentials2016,KatzHandbook2015,DhingraDiseases2014,BessAudiology2003}

Acoustic reflex is based on the fact that a high-intensity acoustic stimulation to one ear causes a middle ear muscle contraction and results in the change in impedance in both ears. A high-intensity tone is delivered to one ear and the reflex is detected from the same or contralateral ear by measuring the change in pressure using a tympanometer.\cite{KramerAudiology2019,GelfandEssentials2016,KatzHandbook2015,DhingraDiseases2014,BessAudiology2003}

Tympanometry and acoustic reflex measurements help to identify the presence of fluid in the middle ear, to evaluate eustachian tube and facial nerve function, to determine the nature of hearing loss, and to assist in the diagnosis of auditory lesion.\cite{KramerAudiology2019,GelfandEssentials2016,KatzHandbook2015,DhingraDiseases2014,BessAudiology2003} 

Figure~\ref{fig-impedance1} shows the interface for impedance audiometry in the application. The interface allows entering tympanometry and acoustic reflex measurement data. As the data is entered, the interface plots the associated tympanogram for both ears.

\subsection{Bithermal Caloric Test Interface}\label{subsec-caloric-interface}
The bithermal caloric test is used to diagnose damage to the acoustic nerve which is involved in hearing and balance. This test induces nystagmus (a vision condition in which the eyes make repetitive, uncontrolled movements) by thermal stimulation of the vestibular system. First, one of the ears is irrigated for 40 seconds alternately with water at 30 \textdegree C and 44 \textdegree C and eyes are observed for the duration from the start of irrigation till the end of nystagmus for both 30 \textdegree C and 44 \textdegree C water irrigations. The process is repeated for the second ear after waiting for 5 minutes. The durations from the start till the end of nystagmus for both ears and for both temperatures are recorded and plotted in a standard graph called calorigram. In addition, two diagnostic metrics are computed from the data measured from this test: canal paresis and directional preponderance.\cite{DhingraDiseases2014}

Canal paresis indicates that the duration of nystagmus for a particular ear canal is less than that from the opposite ear. It is defined as follows:\cite{DhingraDiseases2014,BritishAudiologyCaloric2010}

\begin{equation}
\text{Canal Paresis (\%)}  = \frac{[(R_{44}+R_{30})-(L_{44}+L_{30})]}{[R_{44}+L_{44}+R_{30}+L_{30}]} \times 100 \ \label{eq-canal-paresis} 
\end{equation}

Directional preponderance is a measure of the propensity of the eyes to deviate in one direction over another. It is defined as follows:\cite{DhingraDiseases2014,BritishAudiologyCaloric2010}

\begin{equation}
\text{Directional Preponderance (\%)}  = \frac{[(R_{44}+L_{30})-(L_{44}+R_{30})]}{[R_{44}+L_{44}+R_{30}+L_{30}]} \times 100 \ \label{eq-dir-prepond} 
\end{equation}

where $R_{T}$ and $L_{T}$ are the nystagmus durations for right and left ears, respectively, at a temperature of $T$, for $T=30$ \textdegree C and $44$ \textdegree C.

Figure~\ref{fig-calorigram1} shows the interface for bithermal caloric test. Bithermal caloric test data corresponding to the start and end time of nystagmus for both ears and for both  30 \textdegree C and 44 \textdegree C temperatures can be entered. The interface computes the associated canal paresis and directional preponderance values. A calorigram plot is also generated based on the values entered.

\begin{figure}[t]
\centerline{\includegraphics[scale=1.5]{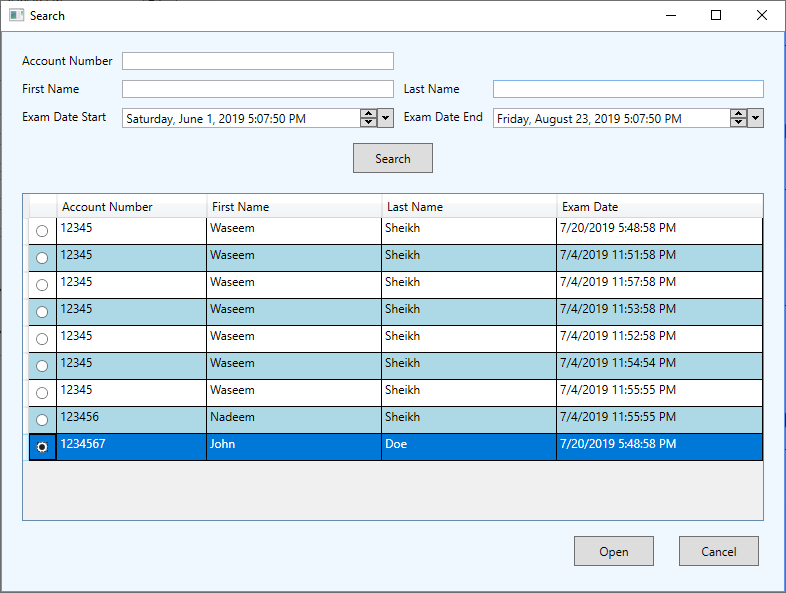}}
\caption{Patient hearing test record search window.\label{fig-search1}}
\end{figure}

\begin{figure}[htbp]
\centering
\begin{minipage}{.5\textwidth}
  \centering
 \includegraphics[scale=0.4]{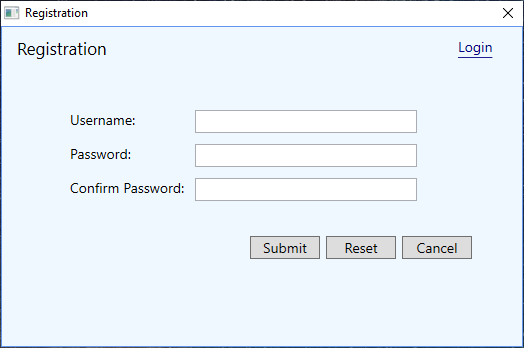}
 \caption{Registration window.\label{fig-register1}}
\end{minipage}%
\begin{minipage}{.5\textwidth}
  \centering
 \includegraphics[scale=0.4]{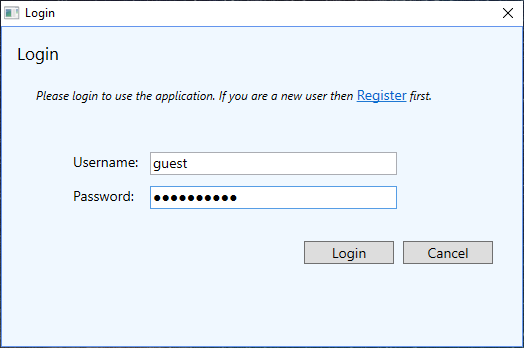}
  \caption{Login window.\label{fig-login1}}
\end{minipage}
\end{figure}

\begin{figure}[htbp]
\centerline{\includegraphics[scale=0.5]{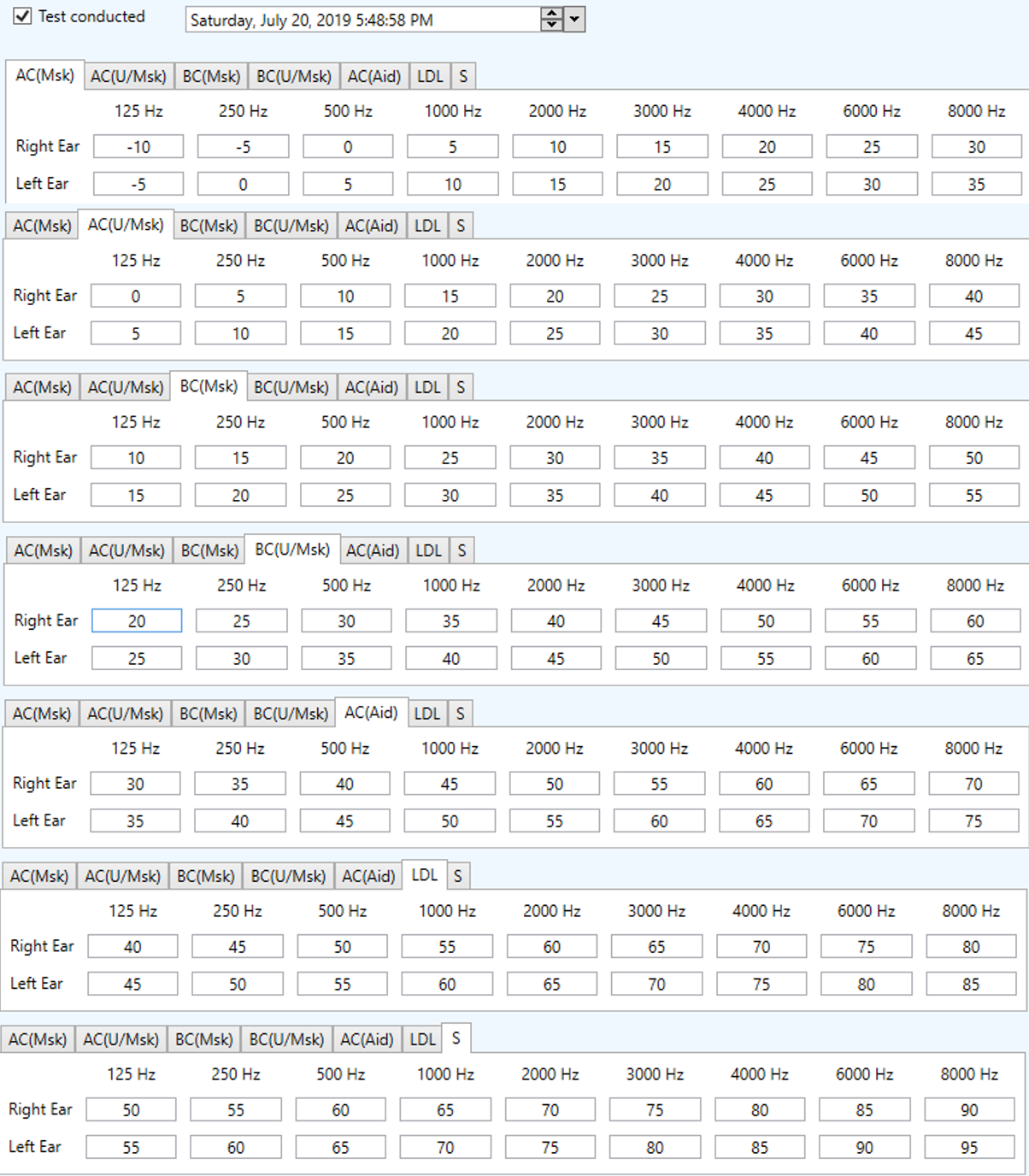}}
\caption{Pure-tone audiometry section of the report.\label{fig-report-puretone}}
\end{figure}

\begin{figure}[htbp]
\centerline{\includegraphics[scale=2]{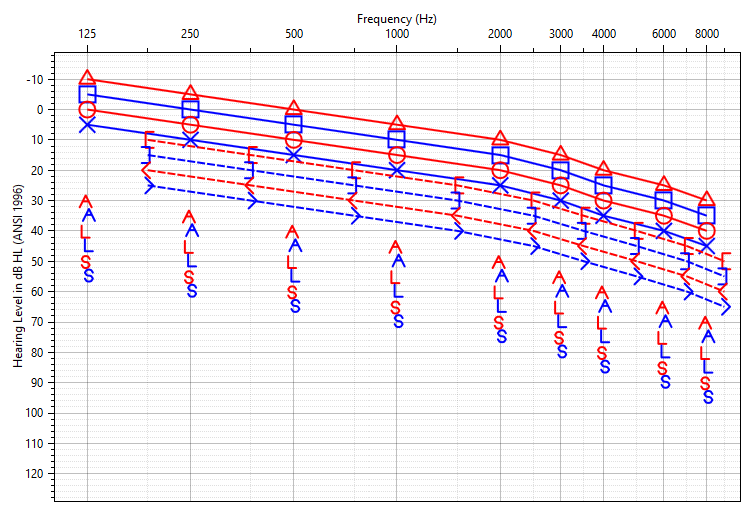}}
\caption{Pure-tone audiogram section of the report.\label{fig-report-audiogram1}}
\end{figure}

\subsection{Other Requirements and Features}\label{subsec-other-features}

For data persistence, the patient's personal and hearing test data are stored in a database. The application also provides the ability to search existing test data and modify it as shown in Figure~\ref{fig-search1}. Standard hearing test data processing algorithms are used to compute test statistics for both ears including average speech perception, hearing impairment, hearing disability, speech recognition threshold (SRT), speech discrimination (SD) score, SD intensity, peak pressure, canal paresis, and directional preponderance.\cite{KramerAudiology2019,GelfandEssentials2016,KatzHandbook2015,DhingraDiseases2014,BessAudiology2003} Standard hearing test graphs are used to plot the hearing test data for visualization purposes. These graphs include pure-tone audiogram (Figure~\ref{fig-puretone1}), speech audiogram (Figure~\ref{fig-speech1}), tympanogram (Figure~\ref{fig-impedance1}), and calorigram (Figure~\ref{fig-calorigram1}). A report of the test data can be generated. This report can be either saved as a PDF file or printed. Sections of this report corresponding to pure-tone audiometry tests are shown in Figures~\ref{fig-report-puretone} and \ref{fig-report-audiogram1}.

A fundamental requirement of the application was that it would be open-source. Consequently, all components and libraries used to build the system are open-source. In addition, the application architecture was designed to make sure that it could easily be extended. The components, framework, libraries used to build the system and the design of the system itself reflect the functional requirement of extensibility. This was to ensure that in addition to using the application right out of the box, the developers will be able to easily extend this application.

Privacy and security were also important functional requirements. Patient and test data are password protected and encrypted in the database. Access to the application is also password protected. Figure~\ref{fig-login1} shows the password-protected login screen for the application. For the first time user, the application provides the ability to register as shown in Figure~\ref{fig-register1}. The username and encrypted password are stored in a registration key on the local machine. Another design requirement was that the application is standalone on a computer and does not reside partly in the cloud. This is to ensure that the application can be used in remote underdeveloped areas where Internet access is not available. The test data stored in the database on a machine can, however, be easily merged with other databases on different machines if such a need arises.

\section{Architecture, Design, and Implementation}\label{sec-design-implementation}

\begin{figure}[htbp]
\centerline{\includegraphics[scale=1.0]{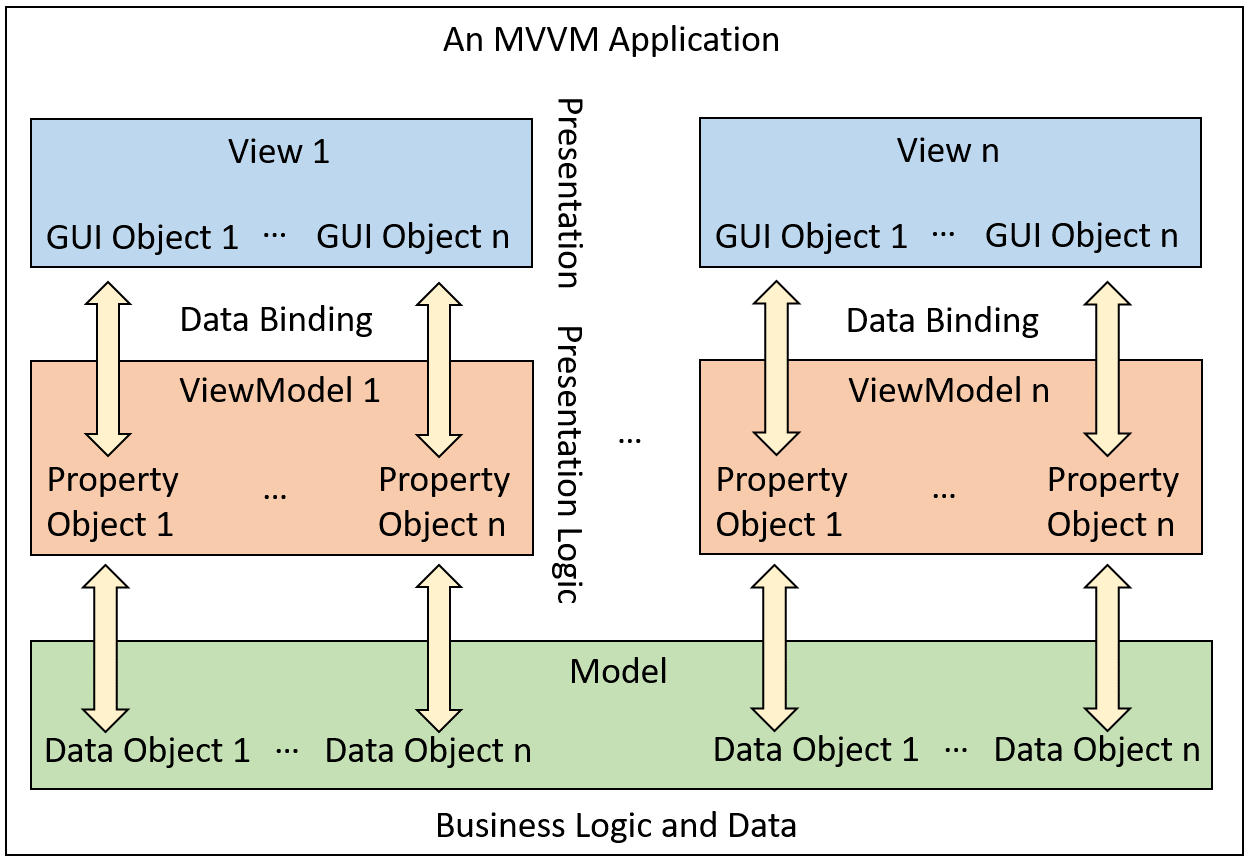}}
\caption{Model-View-ViewModel (MVVM) software architectural pattern.\label{fig-mvvm-pattern}}
\end{figure}

\subsection{Architecture}\label{subsec-architecture}
The main architectural goals of the application were reduced complexity, increased decoupling, faster-streamlined development, ease of extensibility, reusability, ease of maintainability, and ease of testing. In order to achieve these architectural goals, the Model-View-ViewModel (MVVM) software architectural pattern was chosen. MVVM is a variation of the Presentation Model design pattern by Martin Fowler.\cite{FowlerPresentationModel} Another design pattern that preceded and influenced MVVM is the Model-View-Presenter (MVP) design pattern, which is a variation of the Model-View-Controller (MVC) design pattern.\cite{GammaDesign1994} The MVVM pattern was invented by Ken Cooper and Ted Peters to simplify GUI event-driven programming and was later incorporated into Windows Presentation Foundation (WPF), the .NET graphics system by Microsoft\textsuperscript{\textregistered}.\cite{SmithPatterns2009} 

The main idea behind MVVM pattern can be explained from Figure~\ref{fig-mvvm-pattern}. An MVVM application consists of three main components: Model, View, and ViewModel. The application may consist of multiple Views with corresponding ViewModels and Models. The View, which is the XAML (Extensible Application Markup Language) code in WPF, displays a certain layout of data and receives the user's interaction with the View and forwards it to the ViewModel through data binding. Data binding automates the communication between the controls in the View and the Properties in the ViewModel that they are bound to. In the case of WPF, this data binding is provided by XAML. The ViewModel holds a certain shape or state of the data and commands; and binds these data and commands to various controls in the View. The ViewModel acts as a bridge between the Model and the View. It performs all modifications to the data in the Model and exposes a certain shape (using data transformers) or a subset of this data to the View by using Properties to which the XAML code in the View can bind. The Model holds the actual data that the application uses and contains the logic to read it from or write it to persistent storage such as a database. The View depends on the ViewModel but is completely oblivious of the Model. The ViewModel is completely unaware of the View but it depends on the Model. The Model depends on the ViewModel but is completely unaware of the View.

The main advantage of this pattern is that it separates the development of the View from the rest of the application logic thus enabling parallel development of the GUI and application logic in other layers. Other advantages follow from this separation of concerns and include: separate unit testing of the ViewModel and the Model without using the View; redesigning the GUI, by changing the XAML code, without modifying the application logic; reusing the Model and ViewModel layers in different applications; and ease of maintainability.

\begin{figure}[htbp]
\centerline{\includegraphics[scale=1.25]{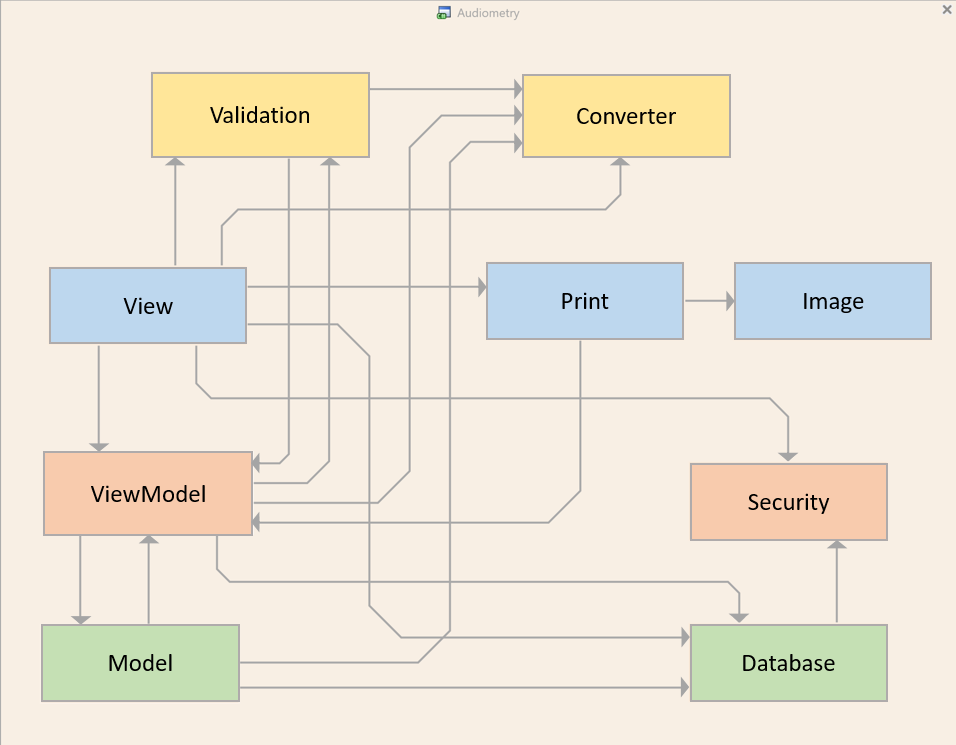}}
\caption{Type dependency diagram of the project namespaces.\label{fig-typedependency-complete}}
\end{figure}

\subsection{Type Dependency View}\label{subsec-type-dependency}

\begin{figure}[t]
\centerline{\includegraphics[scale=1.5]{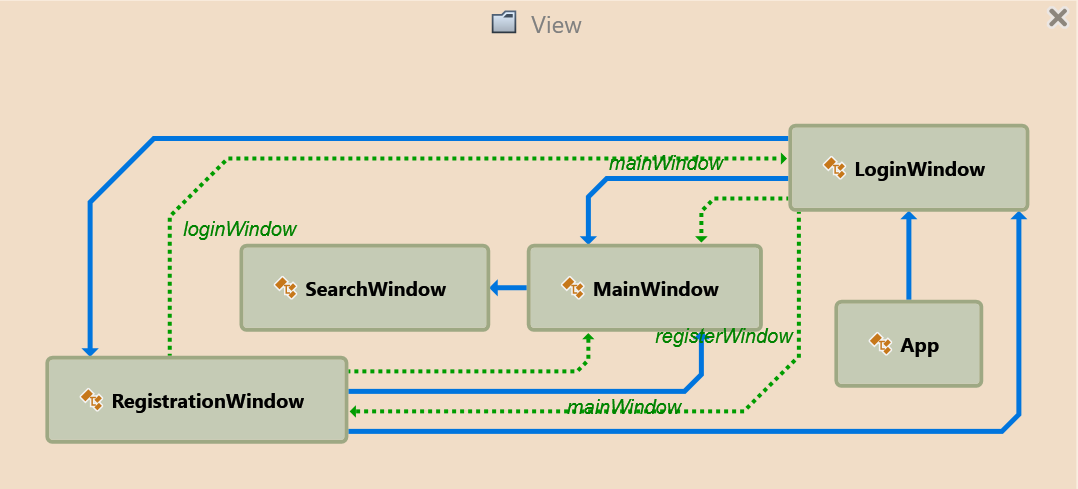}}
\caption{Type dependency diagram of the View (Blue arrows denote usage dependency and dotted green arrows represent private fields).\label{fig-typedependency-View2}}
\end{figure}

Figure~\ref{fig-typedependency-complete} shows the type dependency diagram for the various namespaces as part of the Audiometry application codebase. Each namespace in Figure~\ref{fig-typedependency-complete} contains classes (or types) pertaining to the functionality provided by that namespace. The View namespace contains classes which implement the various views of the application. Almost all of this code is written in XAML with very minimal codebehind in C\#. All of the remaining application logic code is written in C\#. The ViewModel namespace contains classes that implement view models corresponding to the various views of the application. The Model namespace contains classes that contain the actual data that the application consumes and it also provides an ability to read from or write this data to a database. The interaction between View, ViewModel, and Model is the same as explained in Section~\ref{subsec-architecture} for the MVVM pattern. The View depends on the Validation namespace which contains classes that validate the data entered by the user in the GUI. The ViewModel classes also depend on the functionality provided by the Validation namespace to validate the data to and from the GUI.

The Converter namespace contains classes that perform data transformations needed by the View, ViewModel, and Model classes. The Database namespace contains helper classes to read and write from the database. It does not encapsulate the actual database. The View, ViewModel, and Model namespaces depend on the Database namespace. Note that the View and ViewModel do not read from or write to the actual database; only the Model classes do that. The registration view within the View namespace depends on the Database because it has access to the user's password which is used to encrypt the database when a user registers. The ViewModel classes depend on the Database to open and close the database at application launch and exit, respectively. The Model classes read from or write to the actual database tables. The Security namespace contains a class which provides encryption and decryption services to store and validate the user's password in Windows registry. The Database depends on Security because it needs the user's password to open the encrypted database. The Print namespace encapsulates the functionality to print or generate a PDF report of the audiometry test data. The Print namespace depends on the Image namespace which contains classes to render the various GUI components in memory for printing in the report.

The type dependencies among various classes of the namespaces as shown in Figure~\ref{fig-typedependency-complete} reveal the two desired characteristics of a good software architecture: more cohesion and less coupling.

Figure~\ref{fig-typedependency-View2} shows the type interdependencies for the View classes. The View classes consist of RegistrationWindow for registering a user (Figure~\ref{fig-register1}), LoginWindow for logging in a user (Figure~\ref{fig-login1}), SearchWindow for searching patients' test records (Figure~\ref{fig-search1}), and MainWindow for the main application window (Figure~\ref{fig-patient1}). The MainWindow class consists of multiple tabbed Views corresponding to Patient (Figure~\ref{fig-patient1}), Pure-Tone (Figure~\ref{fig-puretone1}), Speech (Figure~\ref{fig-speech1}), Impedance (Figure~\ref{fig-impedance1}), and Bithermal Caloric interfaces (Figure~\ref{fig-calorigram1}). The Pure-Tone View, in turn, consists of three grouped Views including Pure-Tone Audiometry, Special Tests, and Tuning Fork Tests. The Pure-Tone Audiometry View consists of multiple tabbed Views including Air Conduction Masked, Air Conduction Unmasked, Bone Conduction Masked, Bone Conduction Unmasked, Air Conduction Aided, Loudness, and Soundfield. The Special Tests View also consists of multiple tabbed Views including ABLB, SISI, Tone Decay, and Stenger (Figure~\ref{fig-puretone1}). The App launches the LoginWindow which, in turn, can launch a RegistrationWindow for a first-time user, or the MainWindow for an already registered user. The RegistrationWindow launches the MainWindow after successful registration, or it launches the LoginWindow in the case that the user is already registered. The MainWindow can instantiate a SearchWindow if the user wants to search patients' test records.

\begin{figure}[t]
\centerline{\includegraphics[scale=1.0]{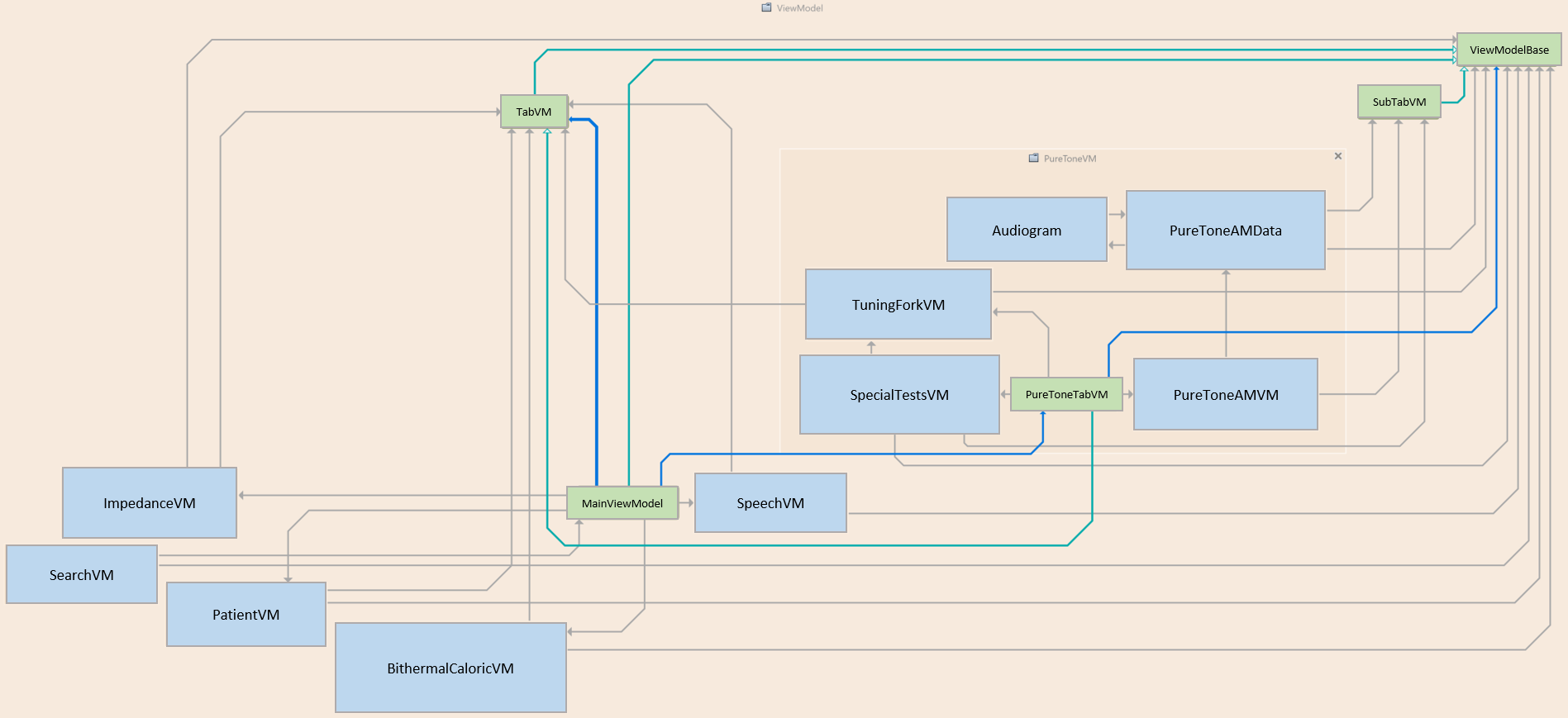}}
\caption{Type dependency diagram of the ViewModel (Blue arrows indicate a usage dependency, green arrows indicate an inheritance relationship, and gray arrows indicate either of the previous relationships).\label{fig-typedependency-ViewModel}}
\end{figure}

\begin{figure}[htbp]
\centerline{\includegraphics[scale=1.0]{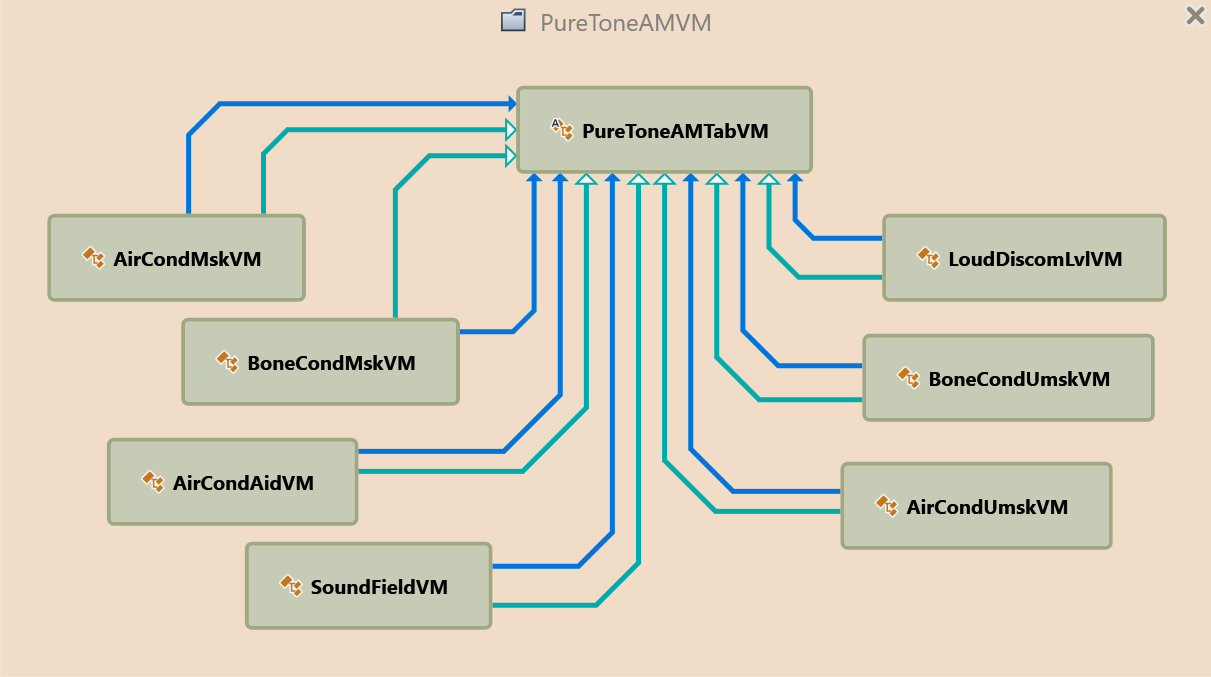}}
\caption{Type dependency diagram of the PureToneAMVM namespace (Blue arrows indicate a usage dependency and green arrows indicate an inheritance relationship).\label{fig-typedependency-PureToneAMVM}}
\end{figure}

The ViewModel type dependency relationships are shown in Figure~\ref{fig-typedependency-ViewModel}. Corresponding to each View class, there is a ViewModel class. The MainViewModel class is a container of all the ViewModel classes. It contains references to PatientTabVM (PatientVM namespace), PureToneTabVM (PureToneVM namespace), SpeechTabVM (SpeechVM namespace), ImpedanceTabVM (ImpedanceVM namespace), and BithermalCaloricTabVM (BithermalCaloricVM namespace) classes. The PureToneTabVM class contains references to AirCondMskVM, AirCondUmskVM, BoneCondMskVM, BoneCondUmskVM, AirCondAidVM, LoudDiscomLvlVM, SoundFieldVM (PureToneAMVM namespace); AblbTabVM, SisiTabVM, ToneDecayTabVM, StengerTabVM (SpecialTestsVM namespace); and TuningForkTestsVM (TuningForkVM namespace) classes. Each PureToneAMTabVM class in PureToneAMVM namespace holds a reference to its corresponding PureToneData class (PureToneAMData namespace). Each PureToneData class contains a reference to its corresponding AudiogramPlot class (Audiogram namespace).

\begin{figure}[t]
\centerline{\includegraphics[scale=1.0]{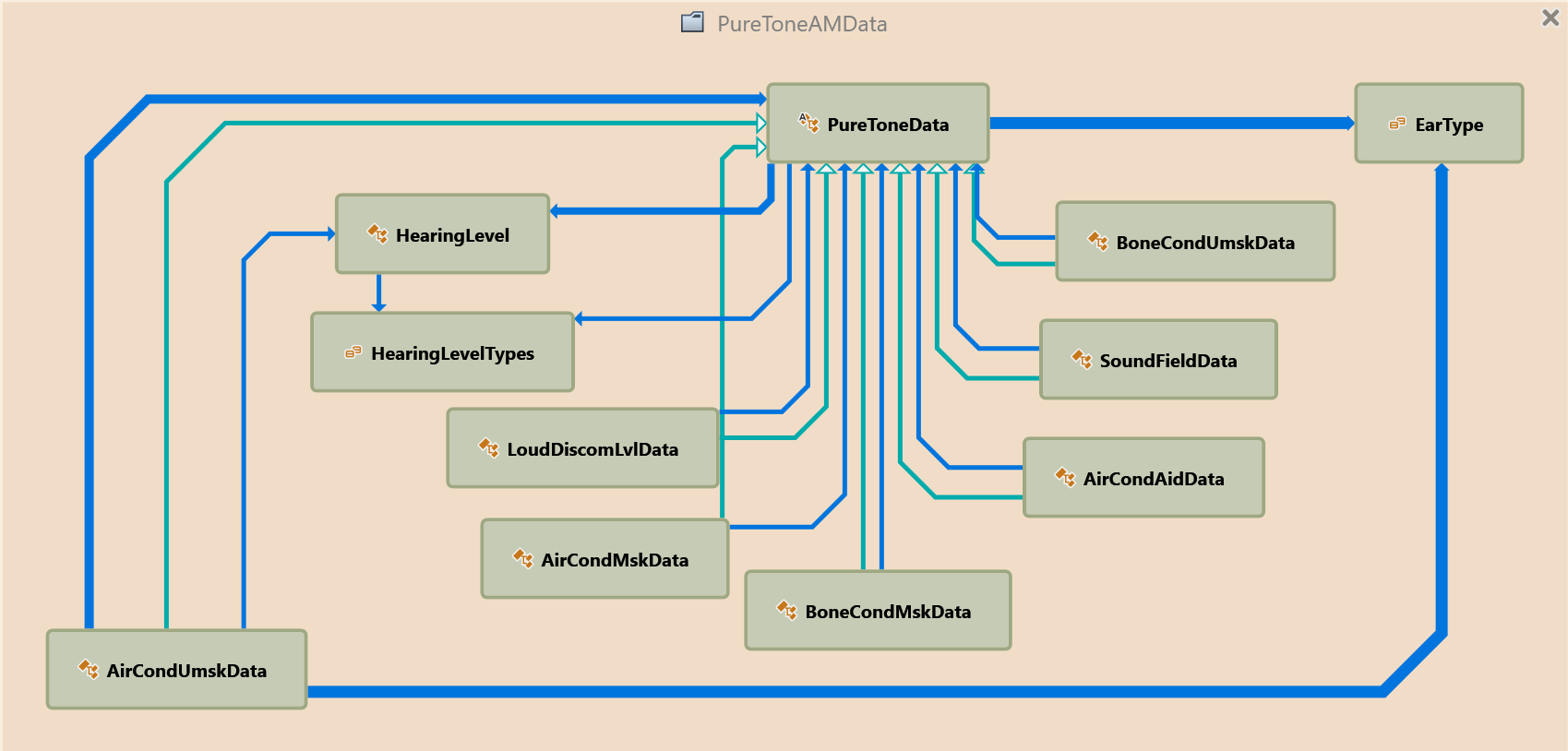}}
\caption{Type dependency diagram of the PureToneAMData namespace (Blue arrows indicate a usage dependency and green arrows indicate an inheritance relationship).\label{fig-typedependency-PureToneAMData}}
\end{figure}

\begin{figure}[htbp]
\centerline{\includegraphics[scale=1.0]{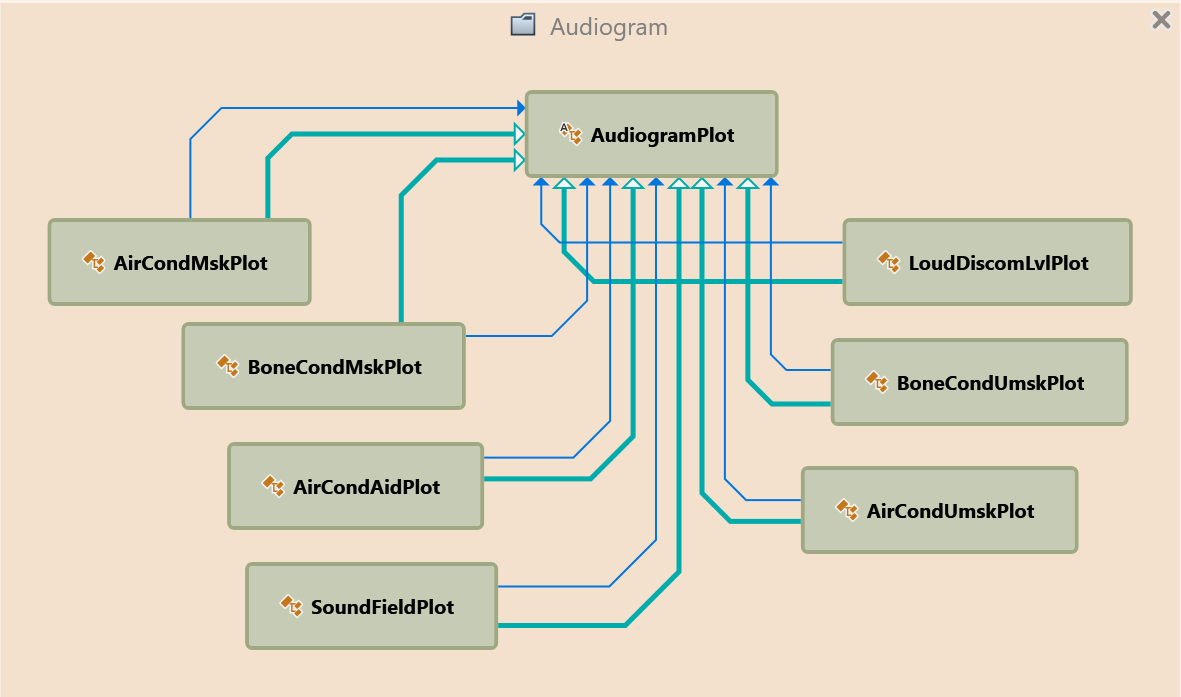}}
\caption{Type dependency diagram of the Audiogram namespace (Blue arrows indicate a usage dependency and green arrows indicate an inheritance relationship).\label{fig-typedependency-PureToneAudiogram}}
\end{figure}

\begin{figure}[t]
\centerline{\includegraphics[scale=0.7]{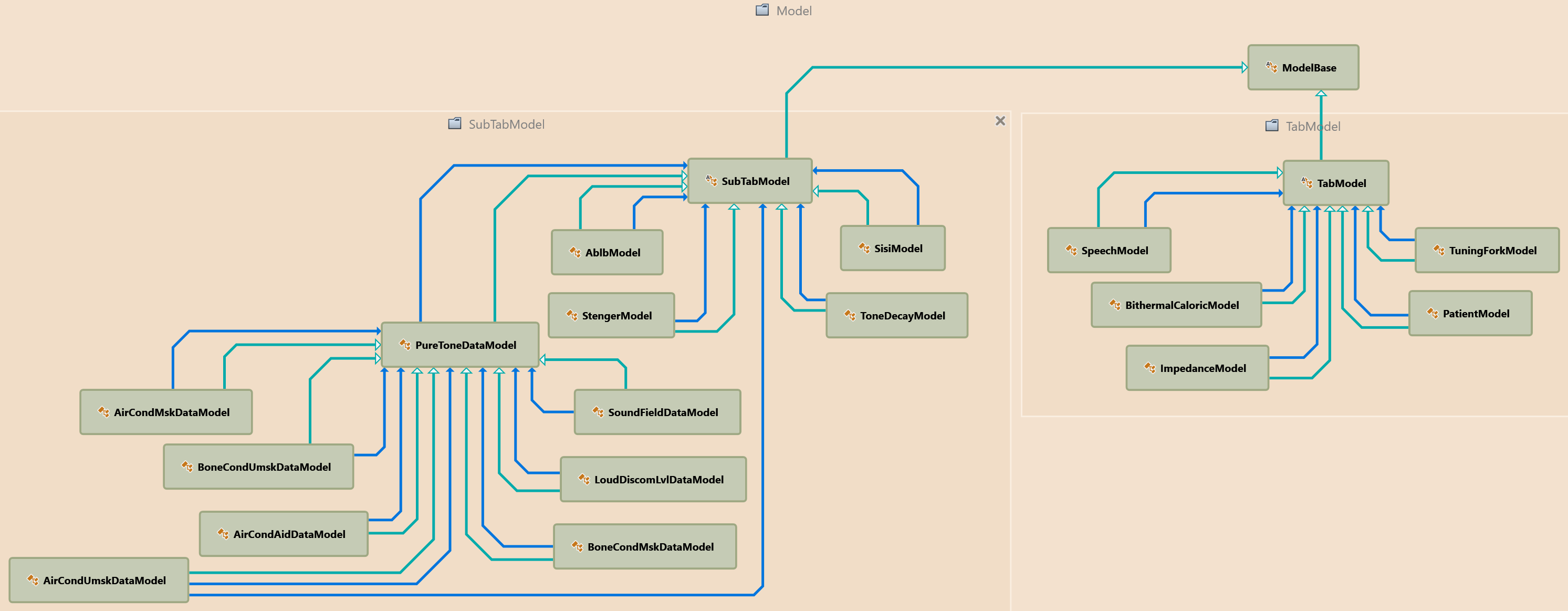}}
 \caption{Type dependency diagram of the Model namespace (Blue arrows indicate a usage dependency and green arrows indicate an inheritance relationship).\label{fig-typedependency-Model}}
\end{figure}

Figure~\ref{fig-typedependency-PureToneAMVM} shows the various classes inside the PureToneAMVM namespace. These contain the pure-tone audiometry ViewModel classes corresponding to air conduction (masked, unmasked, and aided), bone conduction (masked and unmasked), loudness level, and sound field audiometry. Figure~\ref{fig-typedependency-PureToneAMData} shows the various classes inside the PureToneAMData namespace. These classes contain the actual pure-tone audiometry data for air conduction (masked, unmasked, and aided), bone conduction (masked and unmasked), loudness level, and sound field audiometry. Figure~\ref{fig-typedependency-PureToneAudiogram} shows the various classes inside the Audiogram namespace. These classes plot the audiogram curves corresponding to the different types of pure-tone audiometry. Each ViewModel class has a reference to its corresponding Model class. The Model namespace consists of the Model classes shown in Figure~\ref{fig-typedependency-Model}. All Model classes derive from the ModelBase class.

\begin{figure}[htbp]
\centerline{\includegraphics[scale=1.0]{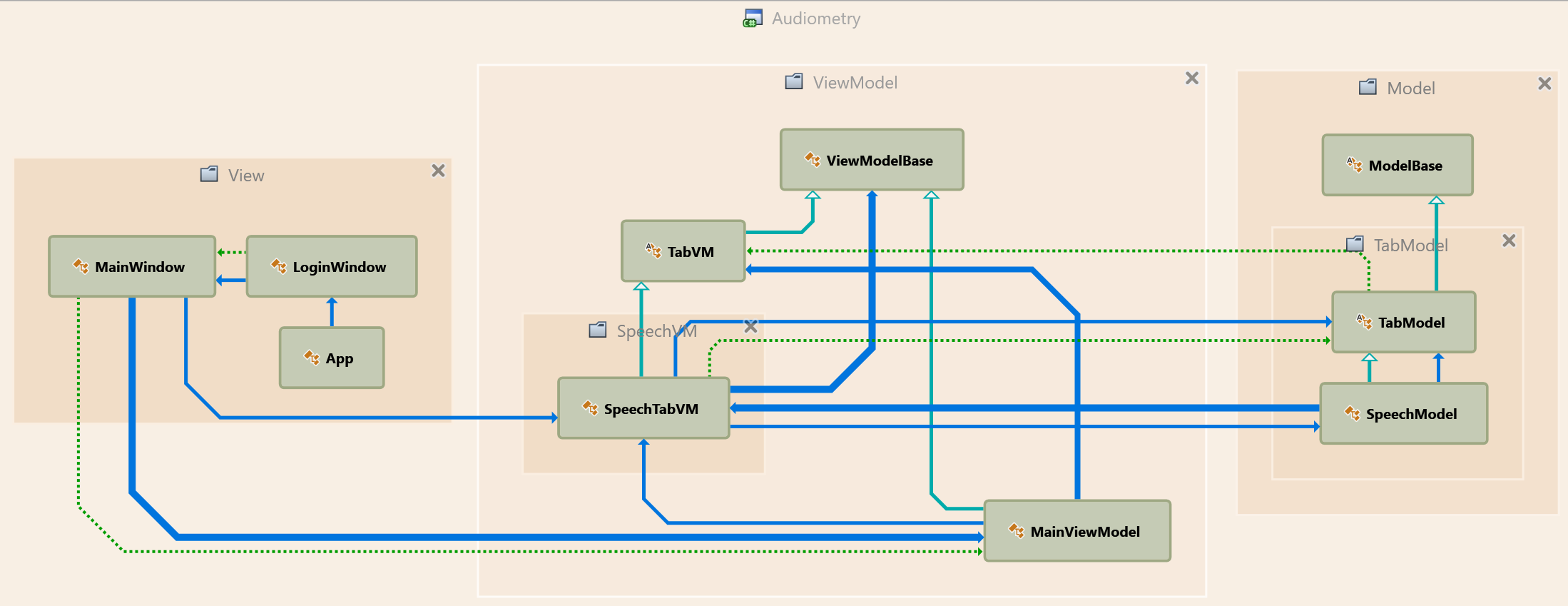}}
\caption{Type dependency diagram of the View, ViewModel, and Model classes corresponding to speech audiometry (Blue arrows indicate a usage dependency, solid green arrows indicate an inheritance relationship, and dotted green arrows indicate a private field).\label{fig-typedependency-SpeechVM-View}}
\end{figure}

\begin{sidewaysfigure}[htbp]
    \includegraphics[width=\textwidth]{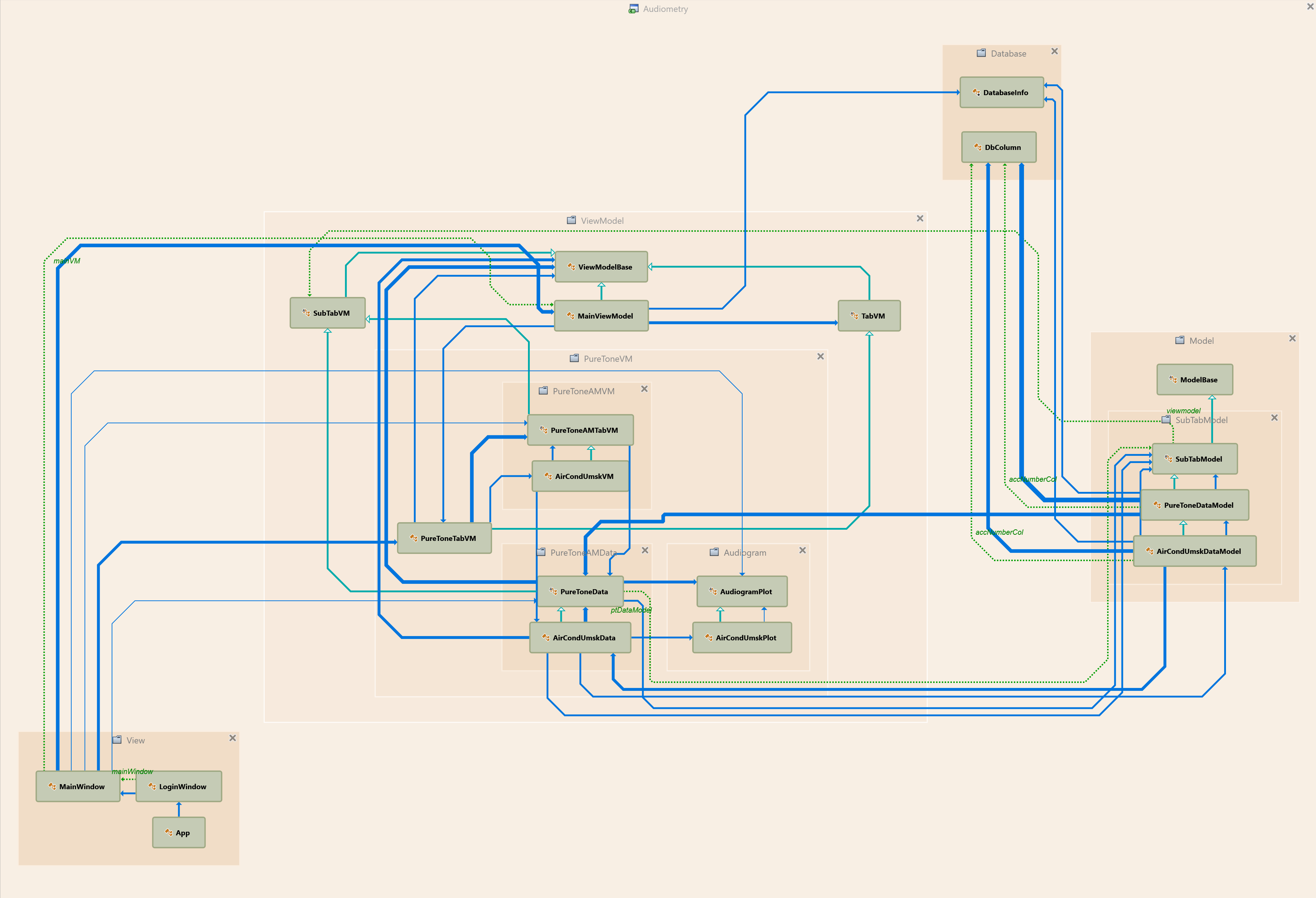}
    \caption{Type dependency diagram of the View, ViewModel, and Model classes corresponding to air conduction unmasked audiometry (Blue arrows indicate a usage dependency, solid green arrows indicate an inheritance relationship, and dotted green arrows indicate a private field).\label{fig-typedependency-AirCondUmsk}}
\end{sidewaysfigure}

The interdependency among a specific View, ViewModel, and Model class can be explained from Figure~\ref{fig-typedependency-SpeechVM-View} which reveals these relationships for the speech audiometry classes. The MainWindow view depends on the MainViewModel which in turn has a reference to SpeechTabVM. The SpeechTabVM depends on SpeechModel and vice versa. Similar relationships among the View, ViewModel, and Model classes for air conduction unmasked audiometry are shown in Figure~\ref{fig-typedependency-AirCondUmsk}. The MainWindow depends on the MainViewModel which has a reference to PureToneTabVM. PureToneTabVM has a reference to AirCondUmskVM. AirCondUmskVM depends on AirCondUmskData which in turn depends on AirCondUmskPlot. AirCondUmskVM also depends on AirCondUmskDataModel. These type interdependencies reveal a systematic and coherent structure of the application framework that lends itself to ease of testing, maintenance, and extension.

\subsection{Class Inheritance View}\label{subsec-class-inheritance}

\begin{figure}[htbp]
\centerline{\includegraphics[scale=1.5]{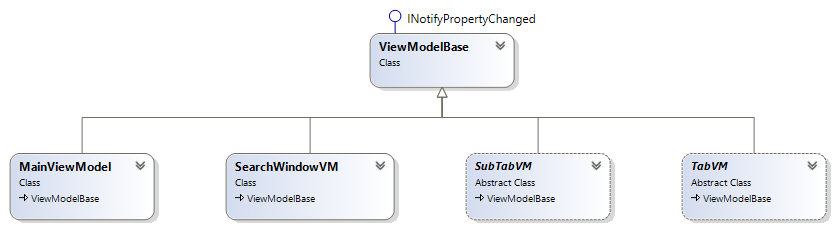}}
\caption{ViewModelBase inheritance diagram.\label{fig-inheritancediagram-ViewModelBase}}
\end{figure}

The class inheritance view also reveals important details about the architecture of the application framework and how to extend it. Inheritance was extensively employed in the project to take advantage of its benefits such as reduced code redundancy, code reusability, manageability, and extensibility. Figure~\ref{fig-inheritancediagram-ViewModelBase} shows the inheritance diagram for the ViewModelBased class which is base class of all ViewModel classes. It implements the INotifyPropertyChanged interface which is used for data binding in WPF. The following classes inherit from the ViewModelBase class: MainViewModel, SearchWindowVM, SubTabVM, and TabVM. The MainViewModel is the ViewModel class for the main application window, SearchWindowVM is the ViewModel class for the search window, SubTabVM is the abstract base class for the ViewModel classes corresponding to subtabs in the main application window, and TabVM is the abstract base class for ViewModels corresponding to tabs in the main application window. The behavior of tabs and subtabs is different and therefore two separate ViewModel base classes were implemented.

\begin{figure}[htbp]
\centerline{\includegraphics[scale=1.5]{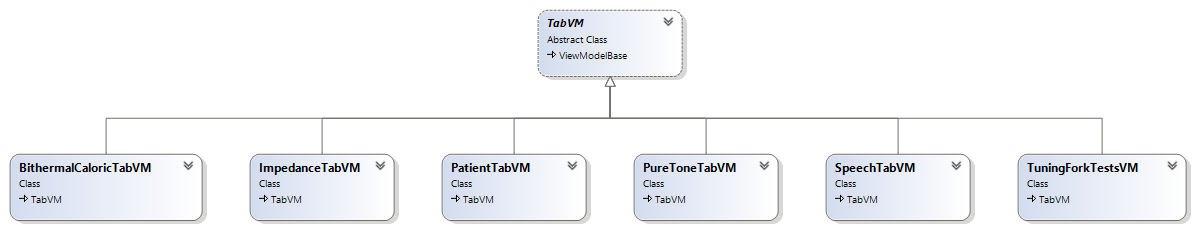}}
\caption{TabVM inheritance diagram.\label{fig-inheritancediagram-TabVM}}
\end{figure}

\begin{figure}[htbp]
\centerline{\includegraphics[scale=1.5]{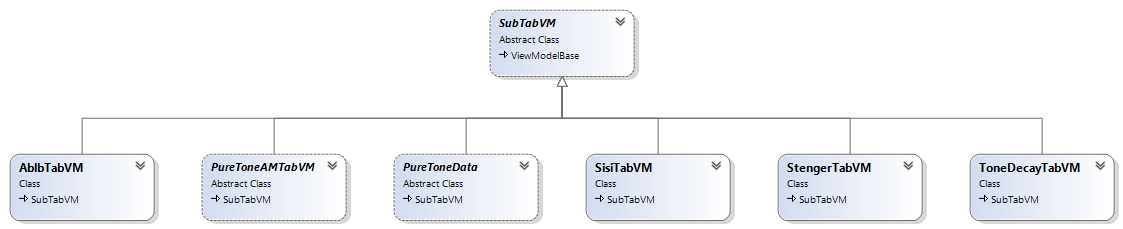}}
\caption{SubTabVM inheritance diagram.\label{fig-inheritancediagram-SubTabVM}}
\end{figure}

\begin{figure}[htbp]
\centerline{\includegraphics[scale=1.5]{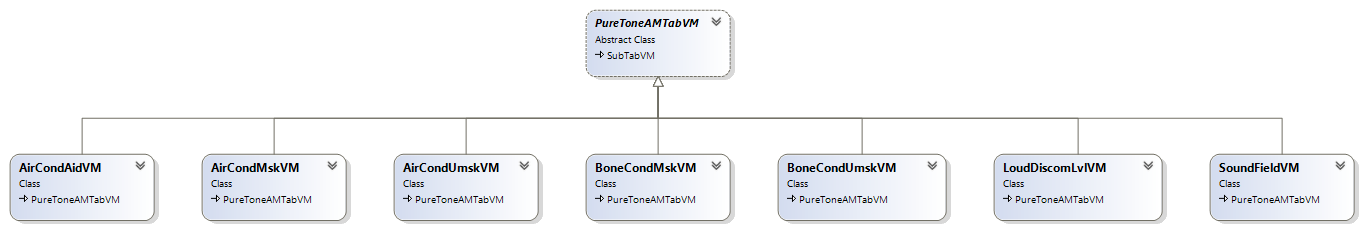}}
\caption{PureToneAMTabVM inheritance diagram.\label{fig-inheritancediagram-PureToneAMTabVM}}
\end{figure}

\begin{figure}[htbp]
\centerline{\includegraphics[scale=1.5]{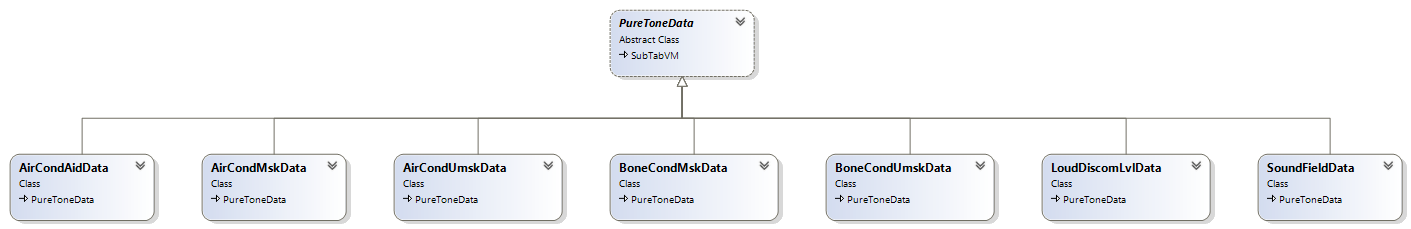}}
\caption{PureToneData inheritance diagram.\label{fig-inheritancediagram-PureToneData}}
\end{figure}

\begin{figure}[htbp]
\centerline{\includegraphics[scale=1.5]{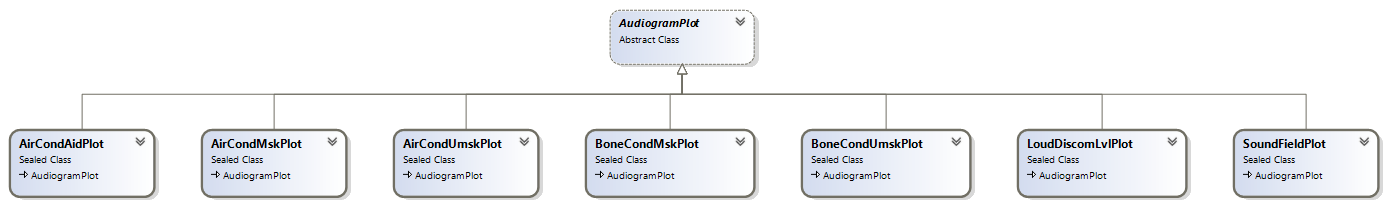}}
\caption{AudiogramPlot inheritance diagram.\label{fig-inheritancediagram-AudiogramPlot}}
\end{figure}

\begin{figure}[t]
\centerline{\includegraphics[scale=1.3]{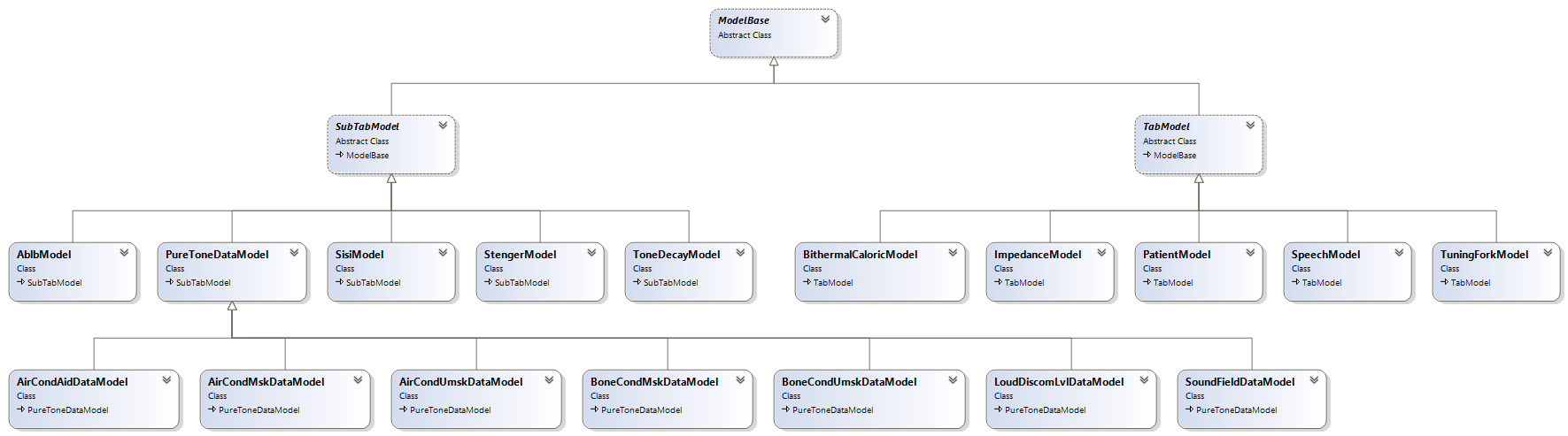}}
\caption{Model inheritance diagram.\label{fig-inheritancediagram-Model}}
\end{figure}

Figure~\ref{fig-inheritancediagram-TabVM} shows the inheritance hierarchy of TabVM class. The following classes inherit from it: BithermalCaloricTabVM, ImpedanceTabVM, PatientTabVM, PureToneTabVM, SpeechTabVM, and TuningForkTestsVM. All of these ViewModel classes correspond to the various tab Views in the main application window. BithermalCaloricTabVM implements the ViewModel for the Bithermal Caloric Test View (Figure~\ref{fig-calorigram1}) , ImpedanceTabVM implements the ViewModel for the Impedance Audiometry View (Figure~\ref{fig-impedance1}), PatientTabVM implements the ViewModel for the Patient Information View (Figure~\ref{fig-patient1}), PureToneTabVM implements the ViewModel for the Pure-Tone Audiometry View (Figure~\ref{fig-puretone1}), SpeechTabVM implements the ViewModel for the Speech Audiometry View (Figure~\ref{fig-speech1}), and TuningForkTestsVM implements the ViewModel for the Tuning Fork Tests View (Figure~\ref{fig-speech1}).

The class inheritance hierarchy for the subtab ViewModel classes is shown in Figure~\ref{fig-inheritancediagram-SubTabVM}. The following subtab ViewModel classes inherit from the SubTabVM abstract base class: AblbTabVM, PureToneAMTabVM, PureToneData, SisiTabVM, StengerTabVM, and ToneDecayTabVM. AblbTabVM is the ViewModel class for the ABLB special test View, PureToneAMTabVM is the abstract base class for all pure-tone audiometry subtab ViewModels, PureToneData is the abstract base class for all pure-tone audiometry data classes, SisiTabVM implements the ViewModel for the SISI special test View, StengerTabVM is the ViewModel class for the Stenger special test View, and ToneDecayTabVM implements the ViewModel class for the ToneDecay special test View.

Figure~\ref{fig-inheritancediagram-PureToneAMTabVM} shows the class inheritance diagram for the pure-tone audiometry ViewModel classes. The following classes derive from the abstract base PureToneAMTabVM class: AirCondAidVM, AirCondMskVM, AirCondUmskVM, BoneCondMskVM, BoneCondUmskVM, LoudDiscomLvlVM, and SoundFieldVM. These classes implement the ViewModels for the Air Conduction Aided, Air Conduction Masked, Air Conduction Unmasked, Bone Conduction Masked, Bone Conduction Unmasked, Loudness Discomfort Level, and Sound Field pure-tone audiometry Views, respectively.

Figure~\ref{fig-inheritancediagram-PureToneData} shows the class inheritance diagram for the pure-tone audiometry data classes. The following classes derive from the PureToneData abstract base class: AirCondAidData, AirCondMskData, AirCondUmskData, BoneCondMskData, BoneCondUmskData, LoudDiscomLvlData, and SoundFieldData. These classes contain the pure-tone audiometry data for the Air Conduction Aided, Air Conduction Masked, Air Conduction Unmasked, Bone Conduction Masked, Bone Conduction Unmasked, Loudness Discomfort Level, and Sound Field Views, respectively.

The class inheritance diagram for the audiogram plot classes is shown in Figure~\ref{fig-inheritancediagram-AudiogramPlot}. The following classes derive from the AudiogramPlot abstract base class: AirCondAidPlot, AirCondMskPlot, AirCondUmskPlot, BoneCondMskPlot, BoneCondUmskPlot, LoudDiscomLvlPlot, and SoundFieldPlot. These classes implement the audiogram plots for the Air Conduction Aided, Air Conduction Masked, Air Conduction Unmasked, Bone Conduction Masked, Bone Conduction Unmasked, Loudness Discomfort Level, and Sound Field Views, respectively.

The complete inheritance hierarchy of the Model classes is shown in Figure~\ref{fig-inheritancediagram-Model}. All Model classes inherit from the ModelBase abstract class. The ModelBase class has two derived classes: SubTabModel and TabModel, both abstract classes corresponding to the Model implementation of a subtab and tab, respectively. The following Model classes inherit from SubTabModel: AblbModel, PureToneDataModel, SisiModel, StengerModel, and ToneDecayModel. These classes implement the Models for ABLB, Pure-Tone Audiometry, SISI, Stenger, and Tone Decay Views. The PureToneDataModel is the parent class for AirCondAidDataModel, AirCondMskDataModel, AirCondUmskDataModel, BoneCondMskDataModel, BoneCondUmskDataModel, LoudDiscomLvlDataModel, and SoundFieldDataModel. These classes implement the Models for Air Conduction Aided, Air Conduction Masked, Air Conduction Unmasked, Bone Conduction Masked, Bone Conduction Unmasked, Loudness Discomfort Level, and Sound Field pure-tone audiometry Views.

The organization of the code into the distinct class hierarchies explained above makes the code reusable, maintainable, testable, and extensible by deriving from the corresponding View, ViewModel, and Model classes.

\subsection{Sequence Diagram View}\label{subsec-sequence-diagram}

\begin{figure}[t]
\centerline{\includegraphics[scale=1.4]{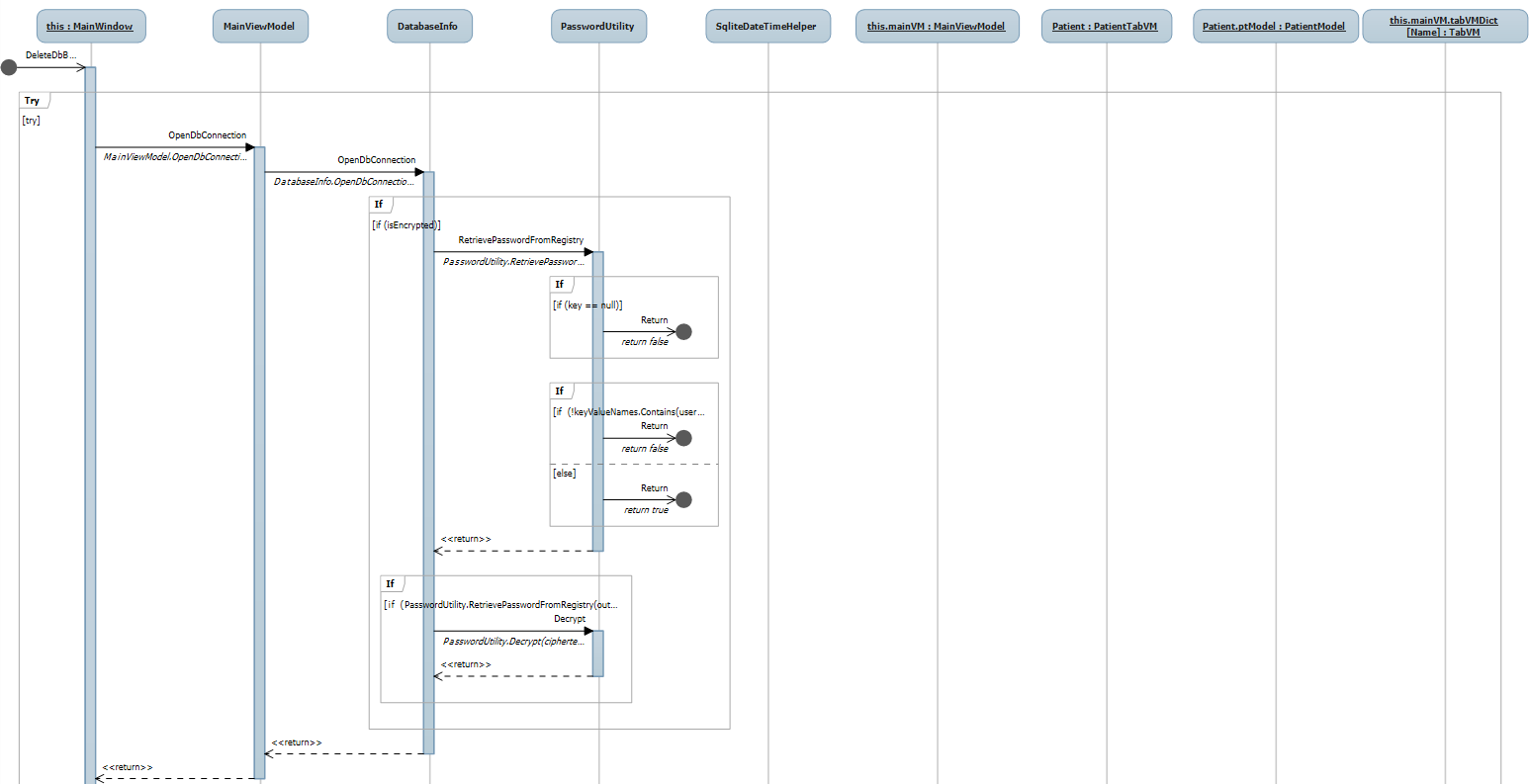}}
\caption{Sequence diagram of DeleteRecordFromDatabase (Part 1).
\label{fig-delete-record-sequence-1}}
\end{figure}

\begin{figure}[htbp]
\centerline{\includegraphics[scale=1.4]{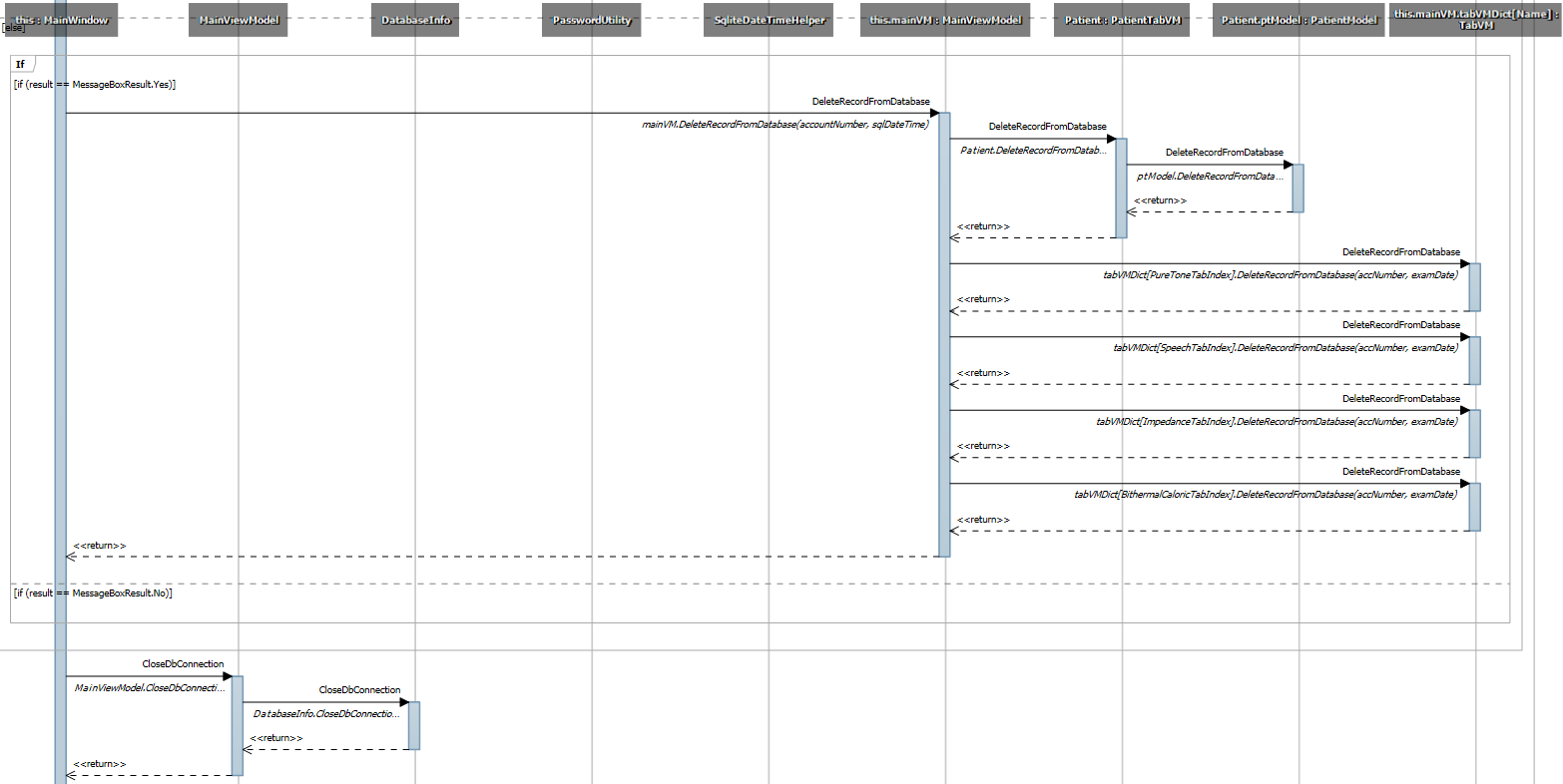}}
\caption{Sequence diagram of DeleteRecordFromDatabase (Part 2).
\label{fig-delete-record-sequence-2}}
\end{figure}

\begin{figure}[htbp]
\centerline{\includegraphics[scale=1.4]{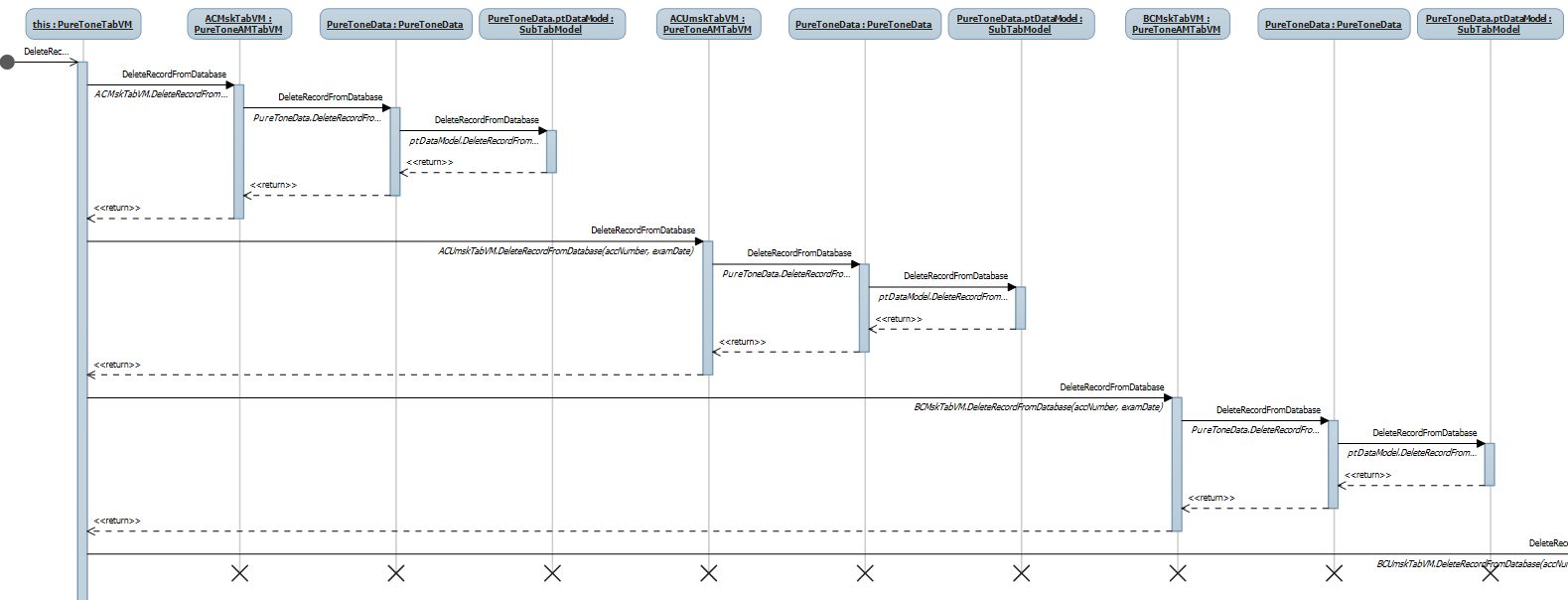}}
\caption{Sequence diagram of DeleteRecordFromDatabase (Part 3).
\label{fig-delete-record-sequence-3}}
\end{figure}

The sequence diagrams of the methods to read, write, modify, or delete a patient's hearing test data reveal important insight into the architecture of the application. These diagrams capture time-focused interaction among various objects in the application to accomplish a specific use case and are helpful for not only understanding the architecture of the application framework but also to extend it. Figures~\ref{fig-delete-record-sequence-1}, \ref{fig-delete-record-sequence-2}, and \ref{fig-delete-record-sequence-3} show the sequence diagram of the DeleteRecordFromDatabase method which deletes a patient's hearing test record from the database. The horizontal axis on the diagrams shows the objects involved in the interaction and the vertical axis represents time progression in the downwards direction. The objects involved in the interaction are listed from left to right in the order of when they first appear in the interaction. The call messages exchanged between different objects are represented by solid arrows while the return messages by dashed arrows. The solid blue rectangle represents the activation of an object, i.e., the duration for which an object is involved in the interaction.

Figure~\ref{fig-delete-record-sequence-1} shows the interaction between MainWindow, MainViewModel, DatabaseInfo, and PasswordUtility objects. Before a record can be deleted, the database has to be opened. In order to open the database, the password is first retrieved from the registry by the PasswordUtility class and then the DatabaseInfo class opens the encrypted password-protected database using this password. Figure~\ref{fig-delete-record-sequence-2} shows the interaction among MainWindow, MainViewModel, PatientTabVM, PatientModel, and tabVMDict objects after the database has been opened in order to delete a patient's hearing test record. The tabVMDict object is a dictionary of the TabVM objects corresponding to the five main tabs of the application, i.e., Patient, Pure-Tone, Speech, Impedance, and Bithermal Caloric. The call to delete a patient's hearing test record is passed on first to MainViewModel object which delegates it to the PatientTabVM object and the remaining four TabVM objects in the tabVMDict dictionary. After the patient's hearing test data is deleted, the connection to the database is closed. Figure~\ref{fig-delete-record-sequence-3} shows part of the sequence diagram for the DeleteRecordFromDatabase method call invoked on the TabVM object for the Pure-Tone Audiometry tab, i.e., PureToneTabVM object. The figure shows the interaction among the PureToneTabVM, AirCondMskVM, AirCondMskData, and AirCondMskDataModel objects. Each pure-tone audiometry Model object is responsible for deleting its pure-tone audiometry data from the database. The DeleteRecordFromDatabase call is delegated from the ViewModel object to the ViewModelData object and finally to the Model object for each kind of pure-tone audiometry. The Model object deletes the particular database table entry corresponding to the patient's audiometry data. This pattern is repeated for all of the pure-tone audiometry objects. Similar sequence diagrams for other use cases of the application reveal important architectural information. This information can be used to make the process of extending the application easier.

\begin{center}
\begin{table}[t]%
\centering
\caption{Database Tables.\label{tab-database}}%
\begin{tabular*}{440pt}{@{\extracolsep\fill}llll@{\extracolsep\fill}}
\toprule
\textbf{No.} & \textbf{Hearing Test Category}  & \textbf{Hearing Test Subcategory}  & \textbf{Database Table Name} \\
\midrule
1  & Patient Information     &                               & PatientInfo       \\
2  & Pure-Tone Audiometry    & Air Conduction Masked         & PureToneACMsk     \\
3  & Pure-Tone Audiometry    & Air Conduction Unmasked       & PureToneACUMsk    \\
4  & Pure-Tone Audiometry    & Bone Conduction Masked        & PureToneBCMsk     \\
5  & Pure-Tone Audiometry    & Bone Conduction Unmasked      & PureToneBCUMsk    \\
6  & Pure-Tone Audiometry    & Air Conduction Aided          & PureToneACAid     \\
7  & Pure-Tone Audiometry    & Loudness Level                & PureToneLDL       \\
8  & Pure-Tone Audiometry    & Sound Field                   & PureToneS         \\
9  & Pure-Tone Audiometry    & Hearing Disability            & HearingDisability \\
10 & Special Tests           & Alternate Binaural Loudness Balance (ABLB)  & Ablb   \\
11 & Special Tests           & Short Increment Sensitivity Index (SISI)    & Sisi   \\
12 & Special Tests           & Tone Decay                    & ToneDecay            \\
13 & Special Tests           & Stenger                       & Stenger              \\
14 & Tuning Fork Tests       &                               & TuningFork           \\
15 & Speech Audiometry       &                               & SpeechAudiometry     \\
16 & Impedance Audiometry    &                               & ImpedanceAudiometry  \\
17 & Bithermal Caloric       &                               & BithermalCaloric     \\
\bottomrule
\end{tabular*}
\begin{tablenotes}
\item Note: The primary key of all tables consists of patient ID and date of exam.
\end{tablenotes}
\end{table}
\end{center}

\subsection{Database Design}\label{subsec-database}

The patients' hearing test data is stored in a local database for persistence. The choice for a local database was made due to the standalone requirements of the application and to keep the design simple. SQLite, an open-source, local, and self-contained relational database management system (RDBMS) was used for this purpose.\cite{Sqlite2019} The data contained in each Model class, pertaining to a specific hearing test, was stored in its own database table. The patient information was stored in a separate table. The list of database tables used by the application is listed in Table~\ref{tab-database}. By mapping each Model class to a database table gives a cleaner design where a change in the table will only propagate to one Model class and vice versa. The Model classes provide the data access layer to the SQLite database tables. The DatabaseInfo class provides helper functions to open and close the database connection. The primary key for all the database tables is the patient ID and the date of the exam. The database is encrypted using the user's password.

\subsection{Security and Privacy}\label{subsec-security-privacy}

For any application storing patients' health-related data, privacy and security is a major concern. One of the requirements of the application was that only authorized users should be allowed access to the application and the patients' personal and test data be encrypted and password-protected. When a user registers the first time, his or her username and encrypted password are stored in the system registry. Future access to the application is only granted if the user's supplied password matches the decrypted password stored in the registry. In addition the database is encrypted and password protected with the user's password, so an unauthorized user would not be able to access the contents of the database. The password encryption and decryption services are provided by the PasswordUtility class.

\subsection{Report Printing}\label{subsec-report-printing}

One of the requirements of the application was that it should be able to print and export a PDF file of the patient's hearing test data. The PrintUtility class provides printing-related services. The various visual controls in the GUI are rendered in memory as bitmap images and then printed or exported as a PDF document. The generated report is organized in accordance with the GUI layout of the application. Figures~\ref{fig-report-puretone} and \ref{fig-report-audiogram1} show the pure-tone audiometry and the pure-tone audiogram sections of the generated report, respectively.

\subsection{Open-Source Components}\label{subsec-components}

The View component of the application framework is written in XAML. The rest of the application is written in C\#. All of the components and libraries used to build the system are open-source. These include WPF, .NET framework, OxyPlot, and SQLite. WPF is a GUI framework by Microsoft\textsuperscript{\textregistered} for creating desktop client applications as part of the .NET framework. OxyPlot is an open-source plot generation library based on the .NET framework.\cite{OxyPlot2019} OxyPlot provides a simple and extensible framework, built upon the Model-View-ViewModel pattern, to generate plots. This library was used for plotting pure-tone audiogram, speech audiogram, impedance audiometry plot, and calorigram plot. SQLite, an open-source library that implements a self-contained, serverless, transactional SQL database engine, was used for persistent storage.\cite{Sqlite2019}

The main reason the application was designed to be targeted for Microsoft\textsuperscript{\textregistered} Windows\textsuperscript{\textregistered} operating systems is due to the fact that an overwhelming majority of computers as part of the healthcare infrastructure in underdeveloped and developing countries run Windows\textsuperscript{\textregistered}. However, the proposed application can also be run on Linux, Unix, Mac OS X\textsuperscript{\textregistered}, and Solaris systems using Mono, which is an open-source cross-platform implementation of Microsoft\textsuperscript{\textregistered}.NET framework.\cite{Mono2019}

\begin{figure}[htbp]
\begin{minipage}{0.25\textwidth}
  \centering
 \includegraphics[scale=1.5]{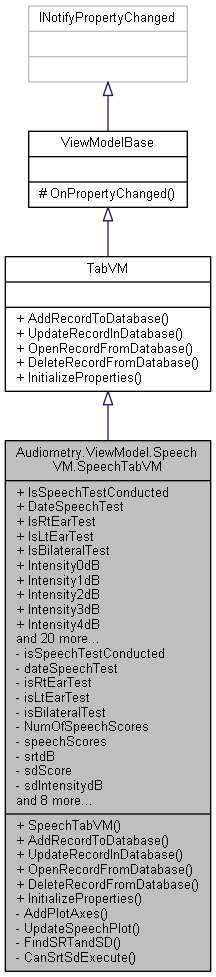}
 \caption{Extending a ViewModel class.\label{fig-extend-SpeechTabVM}}
\end{minipage}%
\begin{minipage}{0.25\textwidth}
  \centering
 \includegraphics[scale=1.5]{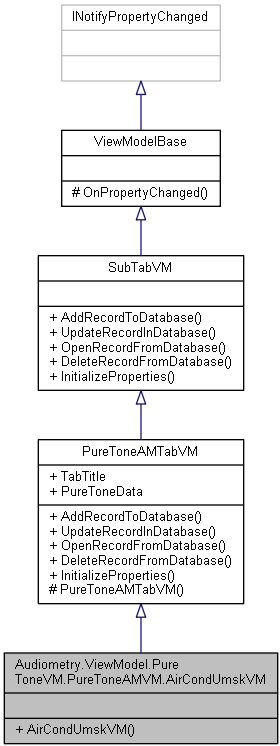}
  \caption{Extending a pure-tone audiometry ViewModel class.\label{fig-extend-AirCondUmskVM}}
\end{minipage}
\begin{minipage}{0.25\textwidth}
  \centering
 \includegraphics[scale=1.5]{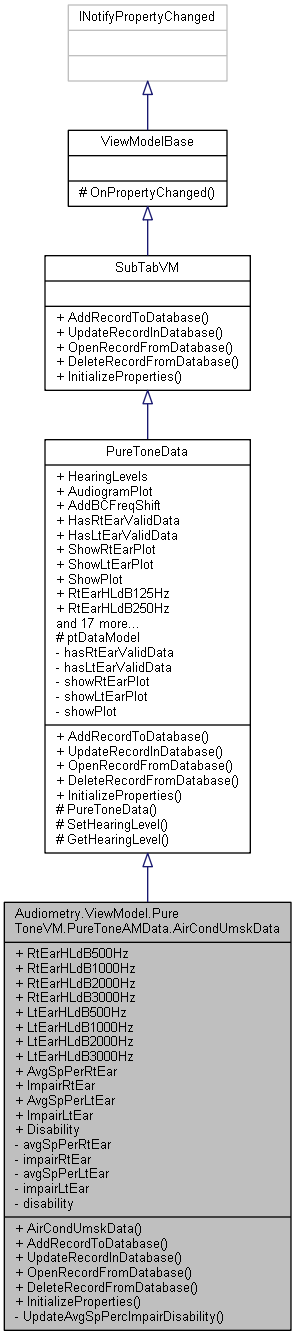}
 \caption{Extending a pure-tone audiometry Data class.\label{fig-extend-AirCondUmskData}}
\end{minipage}%
\begin{minipage}{0.25\textwidth}
  \centering
 \includegraphics[scale=1.5]{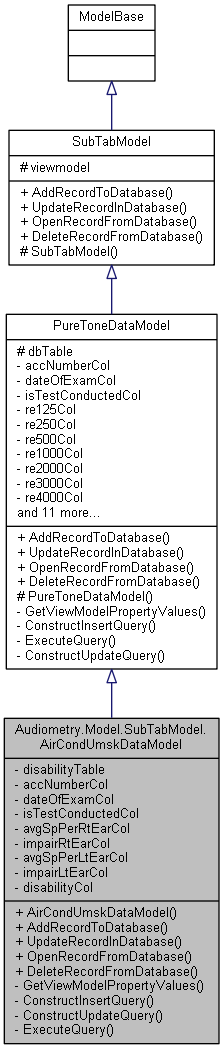}
  \caption{Extending a pure-tone audiometry Data Model class.\label{fig-extend-AirCondUmskDataModel}}
\end{minipage}
\begin{minipage}{0.25\textwidth}
  \centering
 \includegraphics[scale=1.5]{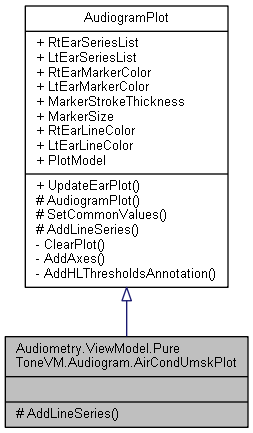}
  \caption{Extending an AudiogramPlot class.\label{fig-extend-AirCondUmskPlot}}
\end{minipage}
\end{figure}

\subsection{Extending the Application}\label{subsec-extend-application}

The proposed application framework was designed to ensure that it would be easily extensible. Consequently, the application framework can be easily extended to create new applications for hearing impairment diagnosis by other researchers and practitioners. Modifications such as adding new hearing tests, removing current ones, or modifying the existing ones are fairly straightforward and systematic due to the architectural choices made. The addition of a new hearing test interface would require adding new View, ViewModel, and Model classes for the new test. In addition, a new database table will have to be created to store the data for the new test. Modifying the GUI of an existing hearing test would require changes to the View, ViewModel, and Model classes for that particular test and the corresponding database table. More specialized versions of the current hearing test interfaces can be created by inheriting (and overriding methods) from the following main classes: tab or subtab View, ViewModel, and Model; pure-tone audiometry ViewModel, pure-tone audiometry Data, pure-tone audiometry Data Model, and pure-tone AudiogramPlot classes as shown in Figures~\ref{fig-extend-SpeechTabVM}, \ref{fig-extend-AirCondUmskVM}, \ref{fig-extend-AirCondUmskData}, \ref{fig-extend-AirCondUmskDataModel}, and \ref{fig-extend-AirCondUmskPlot}.

\section{Practical Experience}\label{sec-practical-experience}

The application was iteratively evaluated and the requirements validated by the coauthor, who is an otorhinolaryngologist and a domain expert in the area of hearing testing, during the development and testing phase of the application. The main criteria used for the evaluation of the application were usability, functionality, accuracy of hearing test data representation and visualization, conformance to the ANSI S3.6-1996 specification,\cite{ANSIS3.6-1996} coverage of a wide variety of hearing tests, ease of use with different hearing test instruments, assistance in diagnosis, ease of managing data, privacy and security, and reduced time for diagnosis.

During the testing and evaluation phase, the domain expert found the application GUI to be very user-friendly. The grouping and arrangement of various test interfaces based on their functional relationships on different tabs were convenient during the patient data collection and diagnosis phases with minimal switching between various interfaces. The hearing test data was easily stored, modified, searched, and deleted. The various diagnostic metrics were automatically computed and the plots automatically generated when the appropriate data was entered without the need for an additional button or menu item clicks. The application in general and the plots generated as part of the application conformed to the ANSI S3.6-1996 specification.\cite{ANSIS3.6-1996} The ability to view different combinations of the audiogram plot curves was particularly helpful in diagnosing certain ear pathologies. In addition, the ability to crossreference different test data and plots was useful for certain hearing impairment diagnosis. A complete hearing test report was easily exported out of the application as a PDF file. In addition, the report was conveniently printed. The application required the user to register the first time by providing a username and password. Afterward, the access to the application was password protected. The database was also encrypted and an unauthorized user could not access the database. The application was found to be responsive in terms of database access, plotting, and data computation. The installation of the application was quite straightforward with just an install of the MSI installer package file.

The proposed application complemented well the clinical evaluation of human hearing dysfunction. It enabled the clinician to reach on a conclusion more methodically, swiftly and accurately. Storing, integrating, and graphically reproducing diverse array of human audio-vestibular digital data for more accurate reporting, later referencing and sharing made the diagnosis of hearing impairments more accurate and time-efficient. 

Many of the ENT, audiological investigation and rehabilitation centers in developing nations, such as the ones where the domain expert worked, still resort to manual hearing test data entry and drawing of audiograms and calorigrams which lacks accuracy, reliability, validity, and reproducibility and is time-inefficient. Introduction of the proposed application aims at helping them to standardize the patients' hearing test data gathering procedure, save consultation time and cost, result in more accurate and better diagnosis, thus promoting better hearing-related health care.

\section{Source Code, Documentation, and Deployment}\label{sec-code-deployment}

The source code for the proposed application framework is publicly available at a GitHub repository.\footnote{Audiometry source code GitHub repository: \url{https://github.com/drwaseemsheikh/audiometry.git}} The repository contains the complete application source code, Doxygen documentation\cite{Doxygen2019}, and WiX (Windows Installer XML) installer source code.\cite{Wix2019} The source code is released under the MIT license. The code was developed using Visual Studio 2017 Community Edition, a freely available IDE.\cite{VisualStudio2019} The WiX installer files were tested on both Windows 7 and Windows 10. By releasing the source code and all the associated documentation, it is our hope that it will foster innovation in the important area of hearing test-related software development and enable other practitioners and researchers to create newer and better hearing test-related applications.

\section{Future Work and Conclusion}\label{sec-conclusion}

Over 5\% of the world's population has disabling hearing loss and a large proportion of it lives in underdeveloped or developing countries with minimal access to hearing testing facilities. Undetected hearing loss not only makes a person less likely to adjust in society and lead a normal life but also incurs significant healthcare and societal costs.

By providing freely available hearing test-related software to medical practitioners in underdeveloped and developing countries, hearing impairment can be more effectively detected at an early stage, thus mitigating its negative effects on a person's life.

Currently, there is a shortage of open-source software for hearing test-related applications. This paper presents the design and implementation of an open-source application framework to assist in the diagnosis of hearing impairment. The proposed application framework is built using the MVVM software architectural pattern which separates the development of GUI from the development of business and back-end logic, thus enabling benefits such as parallel development of GUI, presentation, and business logic, automated unit testing, code reuse, and ease of maintainability and extensibility.

The proposed application can store, process, and visualize data corresponding to tuning fork tests including Weber, Rinne, Schwabach, absolute bone conduction, Teal, and Gelle; speech audiometry; pure-tone audiometry (PTA) including air conduction masked, air conduction unmasked, bone conduction masked, bone conduction unmasked, air conduction aided, loudness level, and sound field; impedance audiometry; bithermal caloric test; and advanced tests including alternate binaural loudness balance (ABLB), short increment sensitivity index (SISI), tone decay, and Stenger.

The application framework is extensible and can be used to develop new hearing test applications by extending the current functionality. The framework is independent of specific hearing test hardware thus making it possible to be used with a wide variety of hearing test hardware. It also provides a unified and uniform interface for storing, processing, and visualizing data from a wide range of hearing tests that traditionally rely on different hardware and software to process and store data.

In addition to the detailed architecture, design, and implementation details of the proposed application framework in this paper, the source code of the application is publicly released to foster innovation in the area of hearing testing software by enabling other practitioners and researchers to extend the current functionality of the proposed application. The source code also contains a WiX installer project which can be used to create MSI installation files for the application to be deployed on a Windows 7 or Windows 10 machine.

The current functionality of the application can be extended and enhanced in various ways. Some important future research directions include adding more hearing impairment diagnostic intelligence into the application, using machine learning and artificial intelligence techniques to increase the accuracy of diagnosis, and a client-server based architecture of the application.


\section*{Acknowledgments}
W. Sheikh would like to thank N. Sheikh for his invaluable insight and expertise in the domain of hearing impairment diagnosis.

\subsection*{Author contributions}

W. Sheikh conceived of developing an open-source application framework to assist in the diagnosis of hearing impairment. N. Sheikh worked as the domain expert and provided all the requirements and details of the GUI design. The software was designed and implemented by W. Sheikh. N. Sheikh also tested and evaluated the software.

\subsection*{Financial disclosure}

None reported.

\subsection*{Conflict of interest}

The authors declare no potential conflict of interests.

\bibliographystyle{unsrt}  
\bibliography{arxiv}  






\end{document}